\documentclass[%
reprint,
superscriptaddress,
amsmath,amssymb,
aps,
prb,
]{revtex4-1}

\usepackage{graphicx}% Include figure files
\usepackage{dcolumn}% Align table columns on decimal point
\usepackage{bm}% bold math
\usepackage{bbm}% bbm fonts
\usepackage{mathrsfs}
\usepackage[table]{xcolor}
\usepackage[colorlinks,citecolor=blue,linkcolor=red,urlcolor=blue]{hyperref}

\DeclareMathOperator{\Tr}{Tr}
\DeclareMathOperator{\sign}{sign}
\DeclareMathOperator{\im}{Im}
\DeclareMathOperator{\re}{Re}

\newcommand*{\dd}{\mathop{}\!\mathrm{d}}
\begin{document}
	
	\preprint{APS/123-QED}

\title{Topological phases induced by charge fluctuations in Majorana wires}
	
\author{M.\,S.\, Shustin}
\affiliation{
\hbox{L. D. Landau Institute for Theoretical Physics, 142432 Chernogolovka, Russia}}
\affiliation{%
Kirensky Institute of Physics, Federal Research Center KSC SB RAS, 660036 Krasnoyarsk, Russia}

\author{S.\,V.\, Aksenov}%
\email{asv86@iph.krasn.ru}
\affiliation{%
Kirensky Institute of Physics, Federal Research Center KSC SB RAS, 660036 Krasnoyarsk, Russia}

\author{I.\,S.\, Burmistrov}
\affiliation{
\hbox{L. D. Landau Institute for Theoretical Physics, 142432 Chernogolovka, Russia}}
\affiliation{Laboratory for Condensed Matter Physics, HSE University, 101000 Moscow, Russia}
	
	\date{\today}% It is always \today, today,
	
	\begin{abstract}

One of the problems concerning topological phases in solid-state systems which still remains urgent is an issue of many-body effects. In this study we address it within perturbative theory framework by considering topological phase transitions related to charge correlations in the extended Kitaev chain model that belongs to the BDI symmetry class. Obtained corrections to a zero-frequency quasiparticle Green's function allow to separate the mean-field and fluctuation contributions to a total winding number. As a result, the phase transitions caused solely by the latter are unveiled. We thoroughly analyze the mechanism of such transitions in terms of fluctuation-induced nodal points and spectrum renormalization. Additionally, features of other quasiparticle properties such as effective mass and damping are discussed in the context of topological phase transitions.

\end{abstract}	
		
\maketitle

%\tableofcontents

\section{\label{sec1}Introduction}

In the last decade, hybrid superconducting (SC) nanowires have been intensively studied. One of the driving forces here is an idea to utilize these structures as a platform for topological quantum computations based on the manipulation of Majorana modes. It was predicted that the nonlocal Majorana states being a hallmark of topologically nontrivial phase might emerge in wires with s-wave SC pairing, spin-orbit coupling and the Zeeman splitting \cite{lutchyn-10,oreg-10}. To detect Majoranas in practice, a semiconducting InSb or InAs wire is used as a core that is covered by a parent SC layer, for example, Al. Such a shell becomes a donor of Cooper pairs in the semiconducting core inducing superconductivity there due to the proximity effect \cite{chang-15}. The latter can be also responsible for the Zeeman splitting if the core is additionally partly coated by a ferromagnetic EuS layer (see, for example \cite{vaitiekenas-21, maiani-21, vaitiekenas-22}). However, an external magnetic field is an alternative option for this purpose \cite{mourik-12,nichele-17,aghaee-22}. 

As a result, the electronic subsystem of the hybrid structure has properties of a spin-polarized superconducting nanowire (SW) with spin-orbit interaction. From the symmetry point of view \cite{zirnbauer-96, altland-97, heinzner-05}, the Hamiltonian modeling such a system belongs to the BDI class, which makes it possible to implement Majorana modes and even many pairs of them \cite{schnyder-08,kitaev-09, turner-11}. An experimental study of such a system led to the discovery of a stable zero-biased peak, which indirectly indicates the Majorana mode presence.  However, the obtained experimental data are interpreted ambiguously and are the subject of discussions (see, for example, \cite{moore-18, reeg-18, prada-20, yu-21}).

There are experimental data indicating that Coulomb interaction in InAs can be controlled by gate electrodes and becomes significant \cite{sato-19}. In particular, for an ensemble of InAs nanowires, the electron transport demonstrated features which can be described in terms of the Luttinger liquid with a constant $K$ corresponding to strong electron correlation regime. 

Noticeable progress has been made in the study of charge-charge interaction in Majorana wires. In specific cases of the Kitaev model \cite{kitaev-01} with the local repulsion, $U$, the twofold degenerate ground states with opposite parity and many-body generalization of the Majorana operators were obtained analytically \cite{katsura-15,miao-17}.
Interestingly, although the interaction 
suppresses the bulk gap in both SW and Kitaev models, the nontrivial phase can be achieved at the lower Zeeman energies and wider range of the chemical potential values (the $U$-proportional renormalizations leading to %that 
such effects can be easily seen already in the mean-field approximation) \cite{gangadharaiah-11,stoudenmire-11,thomale-13}. It was shown that in some models the on-site repulsion itself can be vital for the Majorana modes to emerge \cite{haim-14,thakurathi-18}. 

When the many-body effects are important, Bloch wave functions cannot be used to characterize the spectral and topological characteristics of the system. A natural solution is to apply a Green's function approach \cite{volovik-91}. In particular, topological invariant can be expressed via the Green's function matrix at zero frequency, $\hat{\mathcal{G}}_k(\omega=0)$, with the transitions between phases defined by its zeros and poles \cite{volovik-10,silaev-10,gurarie-11,wang-12,manmana-12,yoshida-14,blason-23}. Furthermore, one can interpret the topological invariant as a winding number of a complex function around the origin (an analogue of the Anderson pseudo vector \cite{anderson-58b}) which real and imaginary parts are the off-diagonal and diagonal components of $\hat{\mathcal{G}}_k(\omega=0)$, respectively \cite{li-18,sugimoto-23}.

As demonstrated in recently studies \cite{choi-23,gavensky-23}, it is possible to decompose the winding number into some bare term corresponding to noninteracting quasiparticles and a part related to quasiparticle's hybridization or interactions. 
Then, one can observe topological phase transitions governed solely by the quasiparticle's interplay. 
In this work we extract the contribution to the topological invariant related to the nearest-neighbour Coulomb repulsion in the extended Kitaev model \cite{vodola-14,patrick-17,alecce-17} beyond the mean-field regime. The corrections to the self-energy obtained up to the second order in perturbation theory make it possible to analyze thoroughly the quasiparticle properties and the mechanism behind the fluctuation-induced phase transitions.

The paper is organized as follows. In Section \ref{sec2} the extended Kitaev model is derived from an SW model in a highly spin-polarized+strongly-correlated regime and its topological phases are considered in the mean-field approximation. In Sec. \ref{sec3} we develop the perturbation theory for the Matsubara normal and anomalous Green's functions up to second order in the intersite repulsive interaction and describe the corresponding effects on quasiparticle properties. In Section \ref{sec4} we show that the topological invariant can be represented as a sum of the mean-field and fluctuation terms. Next, we elucidate in detail how this residual interaction gives rise to the topological phase transitions. 
We end the paper with conclusions (Sec. \ref{sec5}). Some details of computations are delegated to Appendices. Throughout the paper we use unites with $k_B{=}\hbar{=}1$.

\section{Model and Mean-Field Treatment}\label{sec2}

\subsection{Formulation of the model}\label{sec2.1}

We write the SW Hamiltonian describing the electronic system of the hybrid structure in the following form \cite{aksenov-20,shustin-21} 
\begin{eqnarray}
\label{Ham_wire}
\mathcal{H}_{W}  &=&  \frac{1}{2} \sum_{f\sigma} \Big(\xi_{\sigma} a_{f \sigma}^{\dag} a_{f \sigma} + U n_{f \sigma} n_{f \bar{\sigma}} - t_0a_{f \sigma}^{\dag} a_{f+1 \sigma} %+ 
\nonumber \\  &+&\alpha\eta_{\sigma}a_{f \sigma}^{\dag}a_{f+1 \bar{\sigma}}  + 
2\Delta_s \eta_{\sigma}a_{f \sigma} a_{f \bar{\sigma}} %+ 
\nonumber\\
&+&2\Delta_{1s} \eta_{\sigma}\left( a_{f \sigma} a_{f+1 \bar{\sigma}} + a_{f+1 \sigma} a_{f \bar{\sigma}} \right) + \mathrm{h.c.}\Big) %+
\nonumber\\
&+&V_0\sum_{f\sigma\sigma'}n_{f\sigma}\,n_{f+1,\sigma'}.
\end{eqnarray}
Here the terms in the first line represent the 1D Hubbard model with a hopping amplitude $t_0/2$, a spin-dependent single-site energy $\xi_{\sigma} = -\mu + \eta_{\sigma} h $ and a Hubbard repulsive interaction $U$. The Zeeman energy related to the external magnetic field or induced magnetism from the EuS layer is denoted by $h$. The second line of the Hamiltonian contains the terms associated with the spin-orbit coupling with the magnitude $\alpha / 2 $. The proximity-induced $s$-wave superconducting pairing has amplitudes $ \Delta_s $ and $\Delta_{1s}$. The last term in Eq. \eqref{Ham_wire} describes the intersite, $V_0$, Coulomb repulsion of electrons. $a_{f \sigma} (a^{+}_{f \sigma})$ is an annihilation (creation) operator of a fermion at a site $f$ and spin projection $\sigma = {\uparrow, \downarrow}$, $\eta_{\uparrow,\downarrow} = \pm 1$. The electron number operator at the $f$-th site is ${n}_{f \sigma} = a^{+}_{f \sigma} a_{f \sigma}$. Bearing in mind
large $g$-factors in InAs, InSb ($g\simeq 50$) \cite{mourik-12,lutchyn-18} and, consequently, strong Zeeman splitting, in the following it is reasonable to assume that {$ 2|h| \gg |t_0|,\, |\alpha|,\,|\Delta_s|, \,|\Delta_{1s}|$}.

We will focus on the regime of low electron densities and high spin polarization motivated by the available experimental results \cite{vaitiekenas-22, sato-19}. Hence, let us project the Hamiltonian (\ref{Ham_wire}) onto the spin-polarized lower Hubbard band by an operator $$ P = \prod_f \left(X^{00}_f + X^{\uparrow \uparrow}_f \right). $$ The above expression is written using the Hubbard operators acting on single-site states, $ X_{f}^{nm} = | f, \, n \rangle \langle f, \, m | $. In our case there are four states on the site $f$: a state $|f,0\rangle$ with no electrons, two states $|f,\sigma\rangle$ describing one electron with the spin $\sigma$ and a state $|f,2\rangle$ with two electrons possessing opposite spin projections. 

The restriction of Hubbard operators to the low-energy subspace allows to express $\mathcal{H}_W$ in terms of spinless fermion operators:
\begin{eqnarray}
\label{Hub2Ferm}
PX_{f}^{0\uparrow}P=c_{f},~~PX_{f}^{\uparrow\uparrow}P=c^{+}_{f}c_{f} = n_{f}.
\end{eqnarray}
As a result, the desired Hamiltonian in terms of the spinless fermions is given as (see
Appendix \ref{apxA})
\begin{align}
\mathcal{H} =  & \sum_{f=1}^{N}\left(\,\epsilon -\mu\,\right) \, n_{f} \notag \\ -& \frac{1}{2}\sum_{f=1}^{N-1}\Big(t_{1}\,c^{+}_{f}c_{f+1}+ \Delta_{1}\,c^+_{f}c^+_{f+1} - {V}\, n_{f}n_{f+1} + \textrm{h.c.}\,\Big) \notag\\
+& \sum_{f=2}^{N-1}\Big(\,c^{+}_{f+1}\,\hat{t}_{2}\,c_{f-1} +c^+_{f-1}\,\hat{\Delta}_{2}\,c^+_{f+1} + \textrm{h.c.}\,\Big),
\label{Ham_Kit_gen}
\end{align}
where renormalized hoppings and SC pairings are defined as
\begin{equation}
\label{t_delta_gen}
\hat{t}_{2} = t_{2} + F\,n_{f},~~\hat{\Delta}_{2} = \Delta_{2} - G\,n_{f}.
\end{equation}

The Hamiltonian (\ref{Ham_Kit_gen}) is a generalization of the  Kitaev chain Hamiltonian \cite{kitaev-01,lutchyn-10} for the case of long-range hoppings and p-type pairings. The parameters of low-energy model (\ref{Ham_Kit_gen}) depend on the parameters of the original one (\ref{Ham_wire}), see Eqs.\,(\ref{Param})-(\ref{G_eff}). In the regime $1/|\mu| \sim |1/h| \sim 1/\left(U+2|\mu|\right)\ll \max|\{t_0\,,\,\alpha\,,\,\Delta_{s}\,,\,\Delta_{1s} \}|$ the effective model (\ref{Ham_Kit_gen}) describes the low-energy properties of (\ref{Ham_wire}) with a good accuracy. However, in this study we will consider the parameters of (\ref{Ham_Kit_gen}) as independent ones.
This is useful since the effective Hamiltonian similar to (\ref{Ham_Kit_gen}) can be derived from other microscopic models such as the \textit{XY} spin-$1/2$ chain with next-nearest-neighbour frustration \cite{bahovadinov-22}.

The three-center interactions, $\mathcal{H}_3$, 
last line in Eq. \eqref{Ham_Kit_gen}, with amplitudes $F$ and $G$ describe charge-correlated hoppings and SC pairings in the second coordination sphere. The physical meaning of these terms is discussed in Appendix \ref{apxA}, see Eq.\,(\ref{H3_deform}).
Note that the effective interactions are weak, $F,~G \ll 1$, and vanish in the absence of Hubbard repulsion, $U=0$. In that follows, we will treat these interactions at the mean-field level, and the charge fluctuations will be described only by the term $\sim V$. Thus, the Hamiltonian studied further belongs to the BDI symmetry class \cite{karcher-19} and is described by the $\mathbb{Z}_8$ topological classification \cite{fidkowski-11}.

\begin{figure}[htb!]\center
\includegraphics[width=0.4\textwidth]{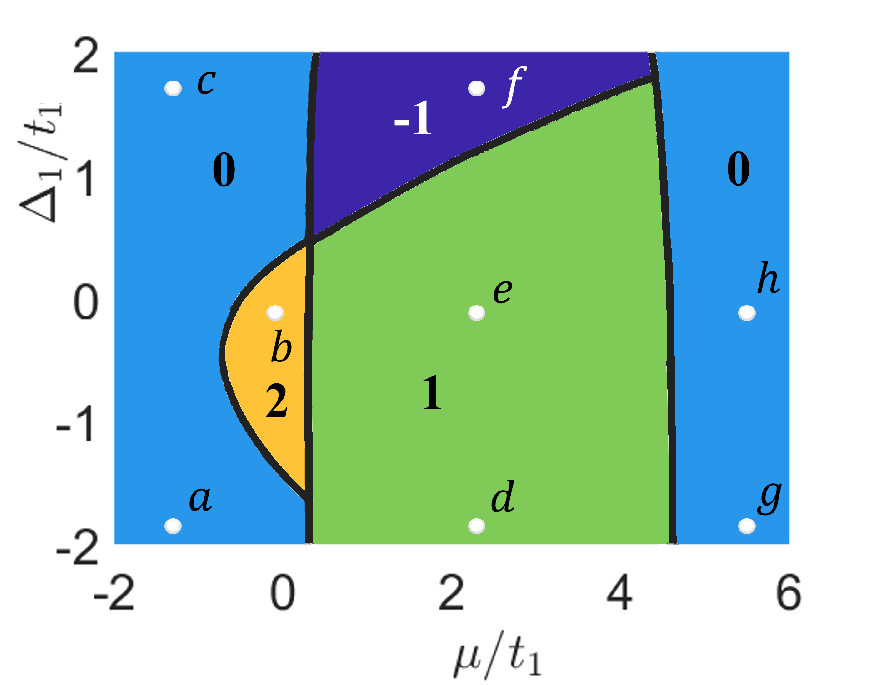}
\caption{The mean-field topological phase diagram of the system. The values of topological index $N_{BDI}^{mf}$ for different parameter ranges
are marked with numbers. When passing through solid black lines, a topological phase transition occurs. 
The integers $-1,0,1,$ and $2$ indicate the topological chatge in a given phase. The white dots, numbered as $a,\dots,h$, correspond to the parameters for which a visualization of $N_{BDI}^{mf}$ values as %an
the winding number of angle $\beta_k = \arctan\left(\tilde{\Delta}_k/\tilde{\xi}_k\right)$ is shown on Fig.\ref{2}. The model parameters in the uniform case in units of $t_1$ are $t_2 = -0.8$, $\Delta_2 = 0.6$, $V=1.2$, $F=0.2$, $G=0.1$.}\label{1}
\end{figure}
%\newpage

\subsection{Mean-field approximation}\label{sec2.2}

At first let us consider the model (\ref{Ham_Kit_gen}) in the mean-field approximation. Using the Hartree-Fock decoupling 
\begin{eqnarray}
c^{+}_{n}c^{+}_{m}c_{f}c_{g} &\to&  \langle c^{+}_{n}c^{+}_{m} \rangle c_{f} c_{g} +
 \langle c_{f}c_{g} \rangle c^{+}_{n}c^{+}_{m} - \langle c^{+}_{n}c_{f} \rangle c^{+}_{m} c_{g} \nonumber\\
 &-& \langle c^{+}_{m}c_{g} \rangle c^{+}_{n}c_{f}+\langle c^{+}_{n}c_{g} \rangle c^{+}_{m} c_{f} +\langle c^{+}_{m}c_{f} \rangle c^+_{n} c_{g},\nonumber
\end{eqnarray}
and introducing the Fermi operators in the momentum space 
\begin{equation}
c_k = \frac{1}{N}\sum_f c_f\,e^{-ikf},
\end{equation}
one can obtain the following  mean-field Hamiltonian in $k$-space from the expression (\ref{Ham_Kit_gen}) \cite{vodola-14,patrick-17,alecce-17}:
\begin{align}
\mathcal{H}_{mf} &= \sum_k\left[\tilde{\xi}_k\,c^+_k\,c_k+\left(\frac{i}{2}\tilde{\Delta}_k\,c_k^+\,c_{-k}^+ - \frac{i}{2}\tilde{\Delta}_k\,c_{-k}\,c_{k}\right)\right], \notag \\
\tilde{\xi}_k & = -\tilde{\mu} -\tilde{t}_1\cos k - \tilde{t}_2\cos 2k , \notag \\
\tilde{\Delta}_k  & = \tilde{\Delta}_1\sin k + \tilde{\Delta}_2\sin 2k .
\label{Ham_mean_field}
\end{align}
Here the renormalized parameters are given as follows
\begin{eqnarray}
\label{xik_Dk}
&& \tilde{\mu} = \mu - 2V\,n - 2F\,\textrm{Re}(\langle c^{+}_{f}c_{f+2} \rangle) - 2G\,\textrm{Re}\langle c_{f}c_{f+2} \rangle, \nonumber\\
&& \tilde{t}_1 = t_1+2V\,\langle c^{+}_{f}c_{f+1} \rangle+ 4F\,\langle c^{+}_{f}c_{f+1} \rangle +4G\,\textrm{Re}\langle c_{f}c_{f+1} \rangle,\nonumber\\
&& \tilde{\Delta}_1 = \Delta_1 + 2V\,\langle c_{f}c_{f+1} \rangle - 4F\,\langle c_{f}c_{f+1} \rangle-4G\,\textrm{Re}\langle c^{+}_{f}c_{f+1} \rangle,\nonumber\\
&&\tilde{t}_2 = t_2 - 2F\,n,~~\tilde{\Delta}_2 = \Delta_2 - G\,n ,  
\end{eqnarray}
and depend on a fermion concentration $n$ as well as on normal and anomalous correlation functions,
\begin{eqnarray}
\label{Av}
&&n = \langle {n}_{f} \rangle = \frac{1}{2} - \frac{1}{4\pi}\int_{-\pi}^{\pi}\frac{\xi_{k}}{\varepsilon_{k}}\,dk\; ,\nonumber\\
&&N_l = \langle c^{+}_{f}c_{f+l} \rangle = -\frac{1}{4\pi}\textrm{Re}{\int_{-\pi}^{\pi}\frac{\xi_k \, e^{ikl}}{\varepsilon_{k}}\,dk}\; , \nonumber\\
&&A_l = \langle c_{f}c_{f+l} \rangle = -\frac{1}{4\pi}\textrm{Im}{\int_{-\pi}^{\pi}\frac{\Delta_k \, e^{ikl}}{\varepsilon_{k}}\,dk}  .
\end{eqnarray}
We emphasize that they should be calculated self-consistently.

The diagonalization of the Hamiltonian can be carried out using the unitary transformation in the Fock space \cite{haken-76}. Denote by $\mathcal{C}$ the Fock space in which the operators $c_k$ act. Consider the unitary rotation to the space $\mathcal{A}$, 
\begin{equation}\label{U_def}
\mathcal{U}\,:\,\mathcal{C} \to \mathcal{A}\,,~~~~~\mathcal{U}^+\,:\,\mathcal{A} \to \mathcal{C},
\end{equation}
in such a way that the Hamiltonian (\ref{Ham_mean_field}) has a diagonal form in the basis $\mathcal{A}$. For this purpose, we define the annihilation operators $\alpha_k$ in the space $\mathcal{A}$ using the relation
\begin{equation}
\alpha_k\,\mathcal{U}\,|\,\Phi\,\rangle = \mathcal{U}\,c_k\,|\,\Phi\,\rangle,~~\mathcal{U}^+\,\alpha_k\,|\,\Psi\,\rangle = c_k\,\mathcal{U}^+\,|\,\Psi\,\rangle,
\end{equation} 
where $|\,\Phi\,\rangle \in \mathcal{C}$ and $|\,\Psi\,\rangle \in \mathcal{A}$. As a result, it becomes possible to define the structure of Fock space in $\mathcal{A}$. 

We set unitary operators $\mathcal{U}$ and $\mathcal{U}^+$ as follows 
\begin{align}
\tilde{\mathcal{U}}= & \prod_{0 \leq k \leq \pi}\exp\left(p_k\,\frac{\beta_k}{2}\left(e^{i\phi_k}\,d_k^{+}d_{-k}^+-e^{-i\phi_k}d_{-k}d_k\right)\right)\nonumber\\
= &\prod_{0 \leq k \leq \pi}\Bigg[1 + p_k\,\left(e^{i\phi_k}\,d_k^{+}d_{-k}^+-e^{-i\phi_k}d_{-k}d_k\right)\,\sin\frac{\beta_k}{2}  \label{U_gen}\nonumber\\
 + & \left(1 - \left(n_k - n_{-k}\right)^2\right)\left(\cos\frac{\beta_k}{2} - 1\right)\Bigg] ,\\
&~~~\mathcal{U} = \tilde{\mathcal{U}}\,|_{d_k \to \alpha_k},~~~\mathcal{U}^+ = \tilde{\mathcal{U}}^+\,|_{d_k \to c_k},\label{U_gen2}
\end{align}
where $n_k = d^+_{k}d_{k}$, $\beta_k\,,\,\phi_k \in \mathbb{R}$ and $p_k=\sign(\tilde{\Delta}_k)$. If we would like to find the result of the action of the operator $\mathcal{U}$ (\,$\mathcal{U}^+$\,) %action 
on the many-body state $|\,\Phi\,\rangle \in \mathcal{C}$ (\,$|\,\Psi\,\rangle \in \mathcal{A}$\,), %that 
then we should act as if the occupation numbers are defined in $\mathcal{A}$ (\,$\mathcal{C}$\,), respectively. It is easy to obtain that
\begin{align}\label{alpha2c}
\alpha_k & = \mathcal{U}\,c_k\,\mathcal{U}^+= \cos\frac{\beta_k}{2}\,c_k - e^{i\phi_k}\,p_k\,\sin\frac{\beta_k}{2}\,c^+_{-k}\,,\\\label{c2alpha}
c_k & = \mathcal{U}^+\,\alpha_k\,\mathcal{U} = \cos\frac{\beta_k}{2}\,\alpha_k +  e^{i\phi_k}\,p_k\,\sin\frac{\beta_k}{2}\,\alpha^+_{-k}\,.
\end{align}
Then, substituting Eq.(\ref{c2alpha})  into the Hamiltonian (\ref{Ham_mean_field}), the requirement for its diagonalization reduces to
\begin{equation}\label{tan_beta}
\tan\beta_k = {\tilde{\Delta}_k}/{\tilde{\xi}_k},~~\phi_k = \pm \pi/2.
\end{equation}
Due to %the 
periodicity of the function $\tan\beta_k$, each value of $\tilde{\Delta}_k/\tilde{\xi}_k$  can correspond to a set of $\beta_k$ values. A natural way to uniquely determine this angle is to fix
\begin{eqnarray}\label{beta_def}
\cos\,\beta_k = \frac{\tilde{\xi}_k}{\varepsilon_k},~~
\sin\,\beta_k = \frac{\tilde{\Delta}_k}{\varepsilon_k},~~
\varepsilon_k = \sqrt{\,\tilde{\xi}^2_k+\tilde{\Delta}^2_k\,}.
\end{eqnarray}
In this case, the $\beta_k$ angle determines the Berry phase of the Bloch states of the mean-field Hamiltonian \eqref{Ham_mean_field}. However, for certain problems, other special choices of $\beta_k$ values that do not violate Eq.\,(\ref{tan_beta}) may be useful. In what follows, in order to construct a perturbation theory with respect to the interaction parameter $V$, it will be convenient to modify the definition of $\beta_k$ in such a way that the unitary operator $\mathcal{U}$ does not perform rotations in the subspaces corresponding to the nodal points of $\tilde{\Delta}_k$:
\begin{eqnarray}\label{c2alpha_NP}
c_{k'} \to \alpha_{k'} = c_{k'},~~\tilde{\Delta}_{k'}=0,~~\beta_{k'} = 0.  
\end{eqnarray}
Consider the action of unitary operator $\mathcal{U}$ on subspace corresponding to the nodal points $k'$. As can be seen from Eq. (\ref{xik_Dk}), the system has symmetric nodal points $\bar{k} = 0,\,\pi$, at which the action of the unitary operator (\ref{U_def}) is reduced to the identity mapping: $\mathcal{U} = \mathcal{I}$. Thus, the conditions of Eq. (\ref{c2alpha_NP}) are performed automatically. However, due to the presence in the model of the next-nearest-neighbor SC pairings additional nodal points emerge,
\begin{equation}\label{k*}
k_* = \arccos\left(-\frac{\tilde{\Delta}_1}{2\tilde{\Delta}_2}\right),~~\textrm{if}~ |\tilde{\Delta}_1|<2|\tilde{\Delta}_2|.
\end{equation}
For these points $k_* \neq - k_*$ (\textrm{mod} $2\pi$) and $\mathcal{U} \neq \mathcal{I}$. So, to satisfy the relations (\ref{alpha2c}) and (\ref{c2alpha_NP}) at these points we will use operator transformations of the following form: 
\begin{eqnarray}\label{uv_def}
&&c_k = u_k\,\alpha_k - i\,v_{k}\,\alpha^+_{-k},~~l_k = \sign\left(\tilde{\xi}_k\right)^{|\sign(\tilde{\Delta}_k)|+1},\\
&&u_k = \sqrt{\frac{1}{2}\left(1+l_k\frac{\tilde{\xi}_k}{\varepsilon_k}\right)},~v_k = \sign(\tilde{\Delta}_k)\,\sqrt{\frac{1}{2}\left(1-l_k\frac{\tilde{\xi}_k}{\varepsilon_k}\right)}.\nonumber
\end{eqnarray} 
Then the Hamiltonian (\ref{Ham_mean_field}) in the space $\mathcal{A}$ takes the form:
\begin{eqnarray}\label{Ham_mean_field_diag}
\mathcal{H}_{mf} &=& \tilde{\xi}_0\,\alpha_0^+\alpha_0 + \tilde{\xi}_{\pi}\,\alpha_{\pi}^+\alpha_{\pi}+\tilde{\xi}_{k_*}\left(\,\alpha_{k_*}^+\alpha_{k_*}+\alpha_{-k_*}^+\alpha_{-k_*}\,\right) \nonumber\\
&+&\sum_{k\neq 0,\pi,k_*}\sqrt{\,\tilde{\xi}^2_k+\tilde{\Delta}^2_k\,}\,\alpha_{k}^+\alpha_{k} + E_0,
\end{eqnarray}
where the terms for which $\tilde{\Delta}_k = 0$ are written in the first line.

The main reason for using transformation (\ref{uv_def}) instead of (\ref{c2alpha}) and (\ref{beta_def}) is that such a choice will allow further study of only particle-like excitations. If we used transformations (\ref{c2alpha}) and (\ref{beta_def}) to diagonalize the Hamiltonian (\ref{Ham_mean_field}), then at the nodal points $k_*$ the operators $\alpha_{k_*}$ would be expressed in terms of the fermion annihilation operators if $\tilde{\xi}_{k_*}>0$ or the hole ones if $\tilde{\xi}_{k_*}<0$,
$$c_{k_*} \to \alpha_{k_*} = \frac{1 + \sign(\tilde{\xi}_{k_*})}{2}
c_{k_*} + \frac{1 - \sign(\tilde{\xi}_{k_*})}{2}
c^+_{-k_*}.$$
Thus, the quasiparticle Green's functions (\ref{G_alpha}), studied in the next section, would describe either particle- or hole-like excitations  (depending on the system parameters), that is inconvenient.

\begin{figure*}[t]\center
\includegraphics[width=0.89\textwidth]{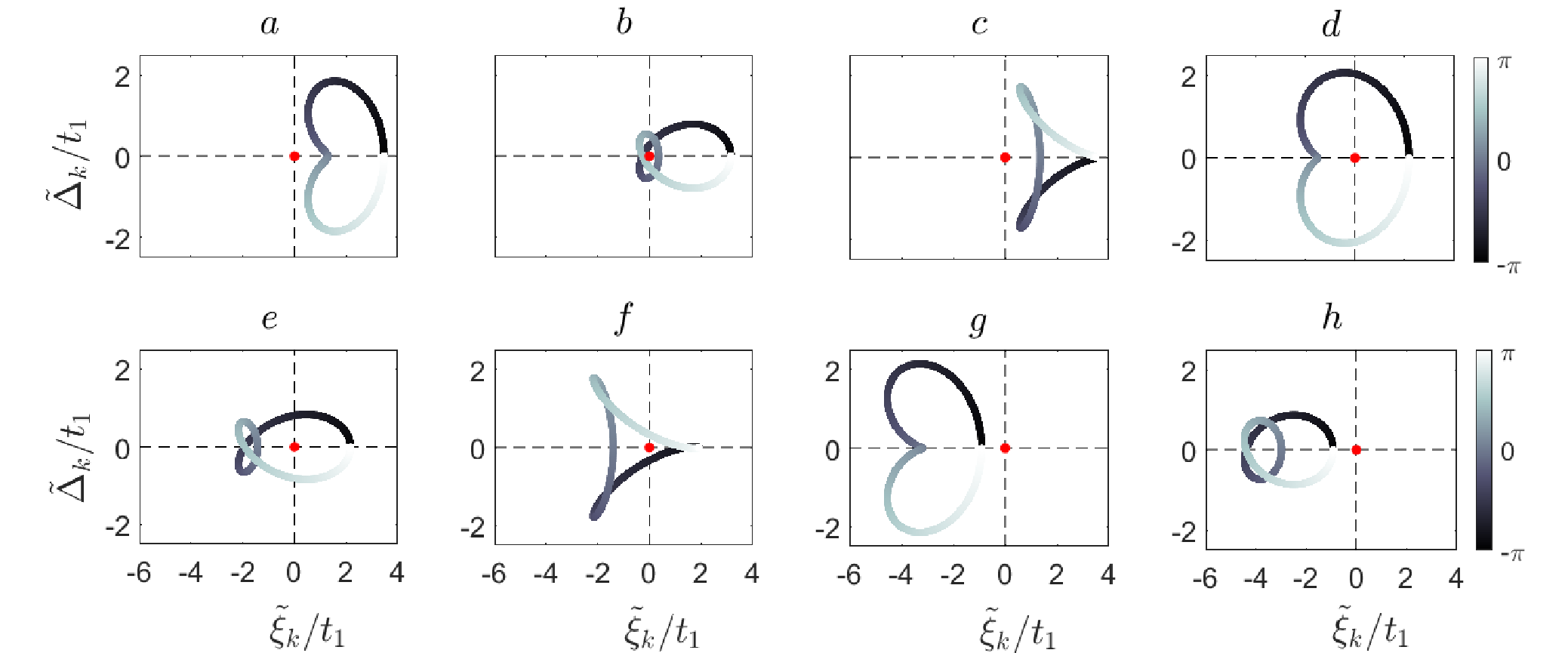}
\caption{Quasimomentum dependencies of renormalized parameters $\tilde{\xi}_k$ and $\tilde{\Delta}_k$ (see Eq.\,\ref{xik_Dk}) for parameters corresponding to the points $a,\dots,h$ on the Fig.\,\ref{1}. Correspondence of color scheme and quasimomentum magnitudes is represented by the gray bar on the right. The origin of coordinates with values $\{\,\tilde{\Delta}_k,\,\tilde{\xi}_k \,\}= \{\,0,\,0\,\}$ is marked with a red dot. The distance from the origin to the curve determines the excitation spectrum of the system $\varepsilon_k = \sqrt{\tilde{\xi}^2_k + \tilde{\Delta}^2_k}$, while the winding number of the curve around the origin in the clockwise direction determines the value of the mean-field topological invariant $N^{mf}_{BDI}$. The modes with $\tilde{\Delta}_k = 0$ and $\tilde{\xi}_k < 0$ are filled in the system GSs. For plots $a,\dots,h$ the same parameters as in Fig. \ref{1} are used. The points 
$a,\dots,h$ in Fig. \ref{1} have the following coordinates in variables $(\,\mu/t_1,\,\tilde{\Delta}_1/t_1\,)$: 
(-1.3, -1.85), (-0.1, -0.12), (-1.3, 1.7), (2.3, -1.85), (2.3, -0.12), (2.3, 1.7), (5.5, -1.85), (5.5, -0.12).}\label{2}
\end{figure*}

Knowledge of the Berry phase $\beta_k$, see Eq.\,(\ref{beta_def}), is useful for the description of the ground-state (GS) structure and topological phase. Then, acting by the unitary operator $\mathcal{U}^+$, Eqs.\,(\ref{U_gen})-(\ref{U_gen2}), on the many-body GS of the Hamiltonian $\mathcal{H}_{mf}$, 
we find that the GS of the system can be one from the following set
\begin{eqnarray}
\label{BCS_func}
&&|\,\Phi_{0}\,\rangle = \prod_{0 \leq k \leq \pi}\left(\,\cos \frac{\beta_k}{2} + i\,p_k^{\,p_k}\,\sin \frac{\beta_k}{2}\,c_k^+c^+_{-k} \right)|\,0\,\rangle_{\mathcal{C}},
\nonumber\\
&&|\,\Phi_{1}\,\rangle = c_0^+|\,\Phi_{0}\,\rangle,~|\,\Phi_{2}\,\rangle = c_{\pi}^+|\,\Phi_{0}\,\rangle,~|\,\Phi_{3}\,\rangle = c_{\pi}^+c_{0}^+|\,\Phi_{0}\,\rangle,\nonumber\\
\end{eqnarray}
where $|\,0\,\rangle_{\mathcal{C}}$ is the vacuum state in the space $\mathcal{C}$. If one of the values of $\tilde{\xi}_0$, $\tilde{\xi}_{\pi}$, $\tilde{\xi}_{k_*}$ is negative, then the corresponding mode will be filled changing the GS \cite{read-00}. The observed effects of such quantum transitions will be described in the next section.

The mean-field topological phase diagram can be obtained by calculating the topological invariant \cite{budich-13,sato-17,valkov-22},
\begin{eqnarray}
\label{N_BDI_MF}
N_{BDI}^{mf} &=& \frac{1}{2\pi}\int_{-\pi}^{\pi}{\varepsilon^{-2}_k}\left(\tilde{\Delta}_k \dot{\tilde{\xi}}_k - \tilde{\xi}_k \dot{\tilde{\Delta}}_k \right)dk\nonumber\\
&=&\frac{1}{2\pi}\oint d\beta_k,
\end{eqnarray}
where $\dot{\tilde{\xi}}_k=\partial_{k}\tilde{\xi}_k$, $\dot{\tilde{\Delta}}_k=\partial_{k}\tilde{\Delta}_k$.
The geometric meaning of the $N_{BDI}^{mf}$ is the winding number of the curve $\{\,\tilde{\xi}_k\,,\,\tilde{\Delta}_k\,\}$, $k \in [-\pi, \pi]$ around the origin. Physically, the $|N_{BDI}^{mf}|$ determines the number of Majorana mode pairs that are localized at the edges of the open and long wire.
A map of $N_{BDI}^{mf}$ values for various magnitudes of the chemical potential $\mu$ and the amplitude of the nearest SC pairings $\Delta_1$ is shown in Fig. \ref{1}. The dots mark the values of the parameters for which the curves $\{\,\tilde{\xi}_k\,,\,\tilde{\Delta}_k\,\}$ are plotted in Fig. \ref{2}. It can be seen that the possibility of realizing phases with several pairs of Majorana modes is related to the existence of nodal points $k_* \neq - k_*$ (\textrm{mod} $2\pi$): their implementation leads to the appearance of an additional loop on the curve $\{\,\tilde{\xi}_k\,,\,\tilde{\Delta}_k\,\}$, which, if it encircles the origin, leads to a change of $N_{BDI}^{mf}$. It can also be seen from Fig. \ref{2} that the conditions for the phases with different $|N_{BDI}^{mf}|$ are determined by the relations:
\begin{eqnarray}
&&|N_{BDI}^{mf}|=0: \sign(\tilde{\xi}_0)=\sign(\tilde{\xi}_\pi) = \sign(\tilde{\xi}_{k_*}),\notag \\
&&|N_{BDI}^{mf}|=1: \sign(\tilde{\xi}_0)=-\sign(\tilde{\xi}_\pi) = \pm\sign(\tilde{\xi}_{k_*}),\label{N_BDI_conditions}\\
&&|N_{BDI}^{mf}|=2: \sign(\tilde{\xi}_0)=\sign(\tilde{\xi}_\pi) = -\sign(\tilde{\xi}_{k_*}).\nonumber
\end{eqnarray}
Note that, in the first two lines (\ref{N_BDI_conditions}), the existence of $k_*$ is not necessary, while for the conditions of the third line it is needed. The boundaries of the topological phases can be carried out by searching the gapless excitations, $\varepsilon_k = 0$. They can be obtained from the conditions $\tilde{\xi}_{0,\pi}=0$ and $\tilde{\xi}_{k_*}=0$:
\begin{multline}\label{mu_bounds}
\tilde{\mu}_{1,2} = 2\left(\tilde{t}_2 \pm \tilde{t}_1 \right),~~~~~~
\tilde{\mu}_* = \frac{\tilde{\Delta}_1}{\tilde{\Delta}_2}\left(\tilde{t}_1 - \tilde{t}_2\,\frac{\tilde{\Delta}_1}{\tilde{\Delta}_2}\right). 
\end{multline}
For $V\,,\,F\,,\,G = 0$ these relations explicitly determine the boundaries of the topological phases. If $V\,,\,F\,,\,G \neq 0$ the above relations turn to nonlinear equations against to $\mu$. 
 
As follows from Fig.\,\ref{2} the realization of quantum phase transitions is not always accompanied by topological quantum transitions. The same topological phases can correspond to different GSs. To clarify this statement let us consider two pair of points on the topological phase diagram Fig.\,\ref{1}. In the first case of `c' and `g' points, we have different types of GSs (with the same fermion parity) as the states with $k=0,~\pi$ either occupied or not. However, the loops in the corresponding Fig.\,\ref{2}c and Fig.\,\ref{2}g are homeomorphic. In the second case of `a' and `b' points, the wave function structure is equivalent (that is $|\,\Phi_0\,\rangle$), but the loops in Figs.\,\ref{2}a and Figs.\,\ref{2}b are already not homeomorphic. Thus, in general situation the topology of the set of wave functions grouped according to the filling of symmetric points ($k=0$ and $k=\pi$) is not equivalent to the topology of the loops group, or mapping $\mathbb{S}^1 \to \mathbb{S}^1$, and the index $N^{mf}_{BDI}$  
describing the degree of the last mapping. 
There is still isomorphism between the Majorana number $\mathcal{M}=(-1)^{N_{BDI}^{mf}}$ and the GS fermion parity \cite{kitaev-01}.

\section{Interaction effects beyond the mean-field approximation}\label{sec3}
\subsection{Self-energy corrections}\label{sec.3.1}

Here we consider the effect of quasiparticle interaction. For simplicity, we will further assume that the three-center interactions with amplitudes $F$ and $G$, renormalize the system parameters only at the mean-field level in accordance with Eq. (\ref{xik_Dk}). To justify this approximation, let us remind that the three-center interactions arise as effective ones for the spin-polarised nanowire in the second order of the operator perturbation theory, see Appendix \ref{apxA} and Eqs. (\ref{F_eff})-(\ref{G_eff}). At the same time, the effective intersite repulsion of fermions, $V$, see Eq.(\ref{V_eff}), results from both renormalizations caused by the virtual transitions to the upper Hubbard subband and the original intersite Coulomb repulsion, $V_0$, which can be significant in the strongly-correlated regime. Then the Hamiltonian can be divided into two parts, the mean-field term and residual ones, $\mathcal{H}=\mathcal{H}_{mf}+\mathcal{H}_{int}$ \cite{coffey-93},
\begin{align}
	\mathcal{H}_{mf} & = \sum_{k} \varepsilon_k\,\alpha_{k}^{+}\alpha_{k}\, \notag \\
	\mathcal{H}_{int} & =\frac{V}{N} \sum_{kpq}\,\Bigr [
\Bigr (A_{k,p,-q;k+p-q}\,\alpha^+_k\,\alpha^+_p\,\alpha^+_{-q}\,\alpha_{k+p-q} \notag \\
	& +\,B_{k,p,-q,-k-p+q}\,\alpha^+_k\,\alpha^+_p\,\alpha^+_{-q}\,\alpha^+_{-k-p+q}\,+ \textrm{h.c.} \Bigr ) \notag \\
 & +C_{k,p;q,k+p-q}\,\alpha^+_k\,\alpha^+_p\,\alpha_{q}\,\alpha_{k+p-q}\Bigl ],\label{H0_H1}
\end{align}
where the amplitudes in $\mathcal{H}_{int}$ have the following structure
\begin{eqnarray}
&&A_{k,p,-q;k+p-q}  = iu_{k}v_{-q}\Bigl \{v_{p}v_{k+p-q}\Bigl [e^{i\left(k-q\right)}-e^{-i\left(k+p\right)} \Bigr ] \nonumber \\ 
&&~~~~~~~~~~~~~~~~~~~~~~~~~~~~+u_{p}u_{k+p-q}\left[e^{-i\left(p-q\right)}-e^{i\left(k-q\right)}\right]\Bigr \},\, \nonumber \\
&&B_{k,p,-q,-k-p+q}  = -u_{k}u_{p}v_{-q}v_{-k-p+q}
  e^{-i\left(p-q\right)},
   \\
&&C_{k,p;q,k+p-q}   {=} v_{k}v_{p}v_{q}v_{k{+}p{-}q}e^{i\left(p{-}q\right)}{+}u_{k}u_{p}u_{q}u_{k{+}p{-}q} e^{i\left(q{-}p\right)}
  %e^{-i\left(p-q\right)} 
  \nonumber
  \\ 
&&~~~~~~~~~~~~~~~+2u_{k}v_{p}v_{q}u_{k+p-q}\left[\cos\left(k+p\right)-\cos\left(p-q\right)\right].\, \nonumber
   \label{ABC}
\end{eqnarray}
From the perturbation theory point of view, the $\mathcal{H}_{int}$ is treated as an interaction operator leading to fluctuation corrections to the mean-field approximation.

Let's introduce the quasiparticle Matsubara Green's function,
\begin{multline}\label{G_alpha}
\hat{\mathcal{G}}^{(\alpha)}_{k}(\tau-\tau')\\
=-\left( \begin{array}{*{20}{c}}
\langle\,T_{\tau}\,\alpha_k(\tau)\,\alpha_k^+(\tau')\,\rangle & \langle\,T_{\tau}\,\alpha_k(\tau)\,\alpha_{-k}(\tau')\,\rangle \\
\langle\,T_{\tau}\,\alpha^+_{-k}(\tau)\,\alpha^+_{k}(\tau')\,\rangle & \langle\,T_{\tau}\,\alpha^+_{-k}(\tau)\,\alpha_{-k}(\tau')\,\rangle
\end{array} \right)\\
=T\sum_{\omega} e^{-i(\tau-\tau')\omega}\,\hat{\mathcal{G}}^{(\alpha)}_{k}(i\omega),\\\hat{\mathcal{G}}^{(\alpha)}_k(i\omega) = \left[\,\hat{g}^{-1}_{k}(i\omega)- \hat{\Sigma}_k(i\omega)\,\right]^{-1},\, \end{multline}
where $\hat{g}_k(i\omega)= i\omega\,\tau_0 - \varepsilon_k\,\tau_z$ is bare propagator of the system depending on $n$th Matsubara frequency, $\omega=\left(2n+1\right)\pi T$. Hereinafter $\tau_{0,x,y,z}$ denote standard Pauli matrices. By definition, the elements of the Fourier transformed matrix are 
\begin{eqnarray}
&&\left[\hat{\mathcal{G}}^{(\alpha)}_{k}(i\omega)\right]_{2,2}=-\left[\hat{\mathcal{G}}^{(\alpha)}_{-k}(-i\omega)\right]_{1,1}^{*},
\nonumber\\
&&\left[\hat{\mathcal{G}}^{(\alpha)}_{k}(i\omega)\right]_{1,2}=-\left[\hat{\mathcal{G}}^{(\alpha)}_{-k}(-i\omega)\right]_{2,1}^{*}.
\end{eqnarray}
The nonzero anomalous components of $\hat{\mathcal{G}}_k(i\omega)$ and the 
self-energy $\hat{\Sigma}_k(i\omega)$ arise from the presence of the anomalous terms with the amplitudes $A_{k,p,-q;k+p-q}$ and $B_{k, p, -q, -k-p+q}$ in $\mathcal{H}_{int}$. However, such anomalous quasiparticle Green's functions have lower order compared to the ones built on the original fermion operators $c_k$. This feature will be used below when calculating the quasiparticle spectrum.

\begin{figure}[htb!]\center
\includegraphics[width=0.45\textwidth]{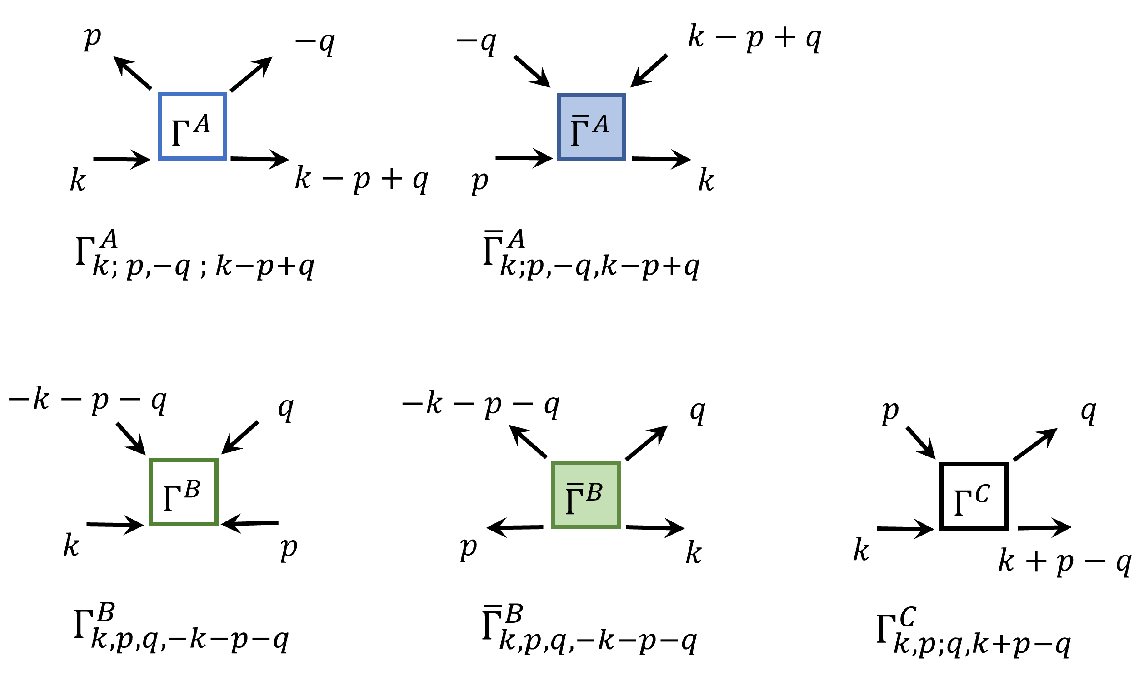}
\caption{Various types of vertices and propagators that appear when calculating self-energy corrections (see Fig.\,\ref{4} and Appendix \ref{apxB}).}\label{3}
\end{figure}

In general, the diagrammatic series generated by $\mathcal{H}_{int}$ \eqref{H0_H1} involve three types of vertices obtained by symmetrizing the interaction amplitudes with respect to the quasimomenta \cite{abrikosov-75},
\begin{eqnarray}
&&\Gamma^{A}_{k;p,-q,k-p+q} =\sum_{P_3}\left(-1\right)^{P_3}\,A_{k;P_3\{p,-q,k-p+q\}}\,,\nonumber\\
&&\Gamma^{B}_{k,p,q,-k-p-q} =\sum_{P_4}\left(-1\right)^{P_4}\,B_{P_4\{k,p,q,-k-p-q\}}\,,\label{ABC_vortices} \\
&&\Gamma^{C}_{k,p;q,k+p-q} =\sum_{P_2,\,P'_2}(-1)^{P_2 + P'_2}\,C_{P_2\{k,\,p\}P_2'\{q,\,k+p-q\}}\,.\nonumber
\end{eqnarray}
In the above expressions all possible permutations in a set of two, three and four quasimomenta, $\{k,p\}$ ($\{q,k+p-q\}$), $\{p,-q,k-p+q\}$ and $\{k,p,q,-k-p-q\}$ are denoted by $P_2$ ($P'_2$), $P_3$ and $P_4$, respectively. In turn, $(-1)^{P_{2,3,4}}$ provides the parity of such permutations. The quasimomentum variables under symmetrization should correspond to propagators that have the same direction with respect to the vertices, i.e. either entering or leaving them (see Fig.~\ref{3}).

\begin{figure}[t]\center
\includegraphics[width=0.5\textwidth]{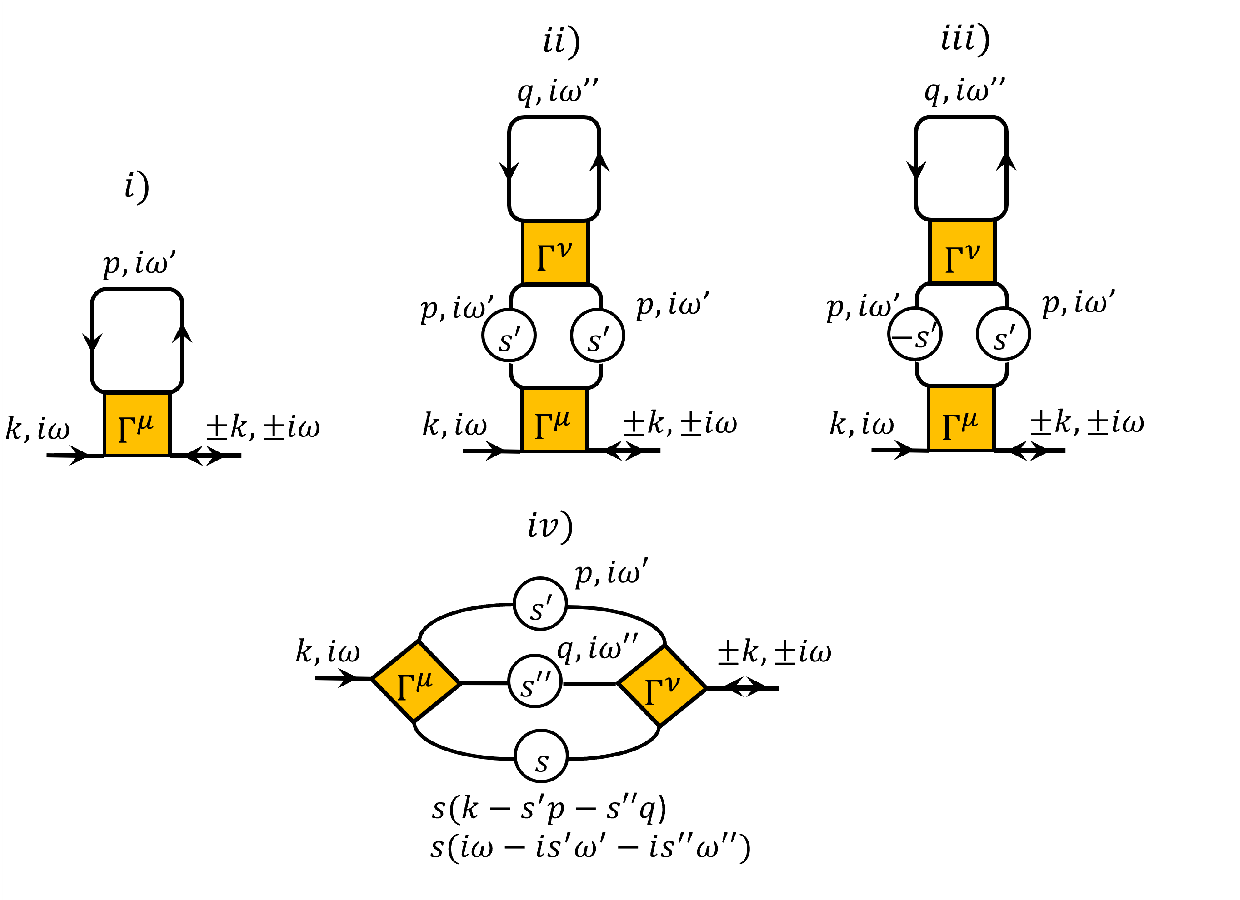}
\caption{Four classes of irreducible diagrams providing the first- and second-order corrections to the self-energy. Variables $s,\,s',\,s'' = \pm 1$ in the circles on the propagator lines determine the direction of their arrows. If $s,\,s',\,s'' = 1$ ($s,\,s',\,s'' = -1$), then the circles have to be replaced by arrows directed from $\Gamma^{\mu}$ to $\Gamma^{\nu}$ (from $\Gamma^{\nu}$ to $\Gamma^{\mu}$) vertices. }\label{4}
\end{figure}

Expanding the scattering matrix up to the second order in $\mathcal{H}_{int}$ and averaging according to Wick's theorem, we obtain 15 irreducible diagrams describing the corrections for the normal and anomalous  Green's functions. The explicit form of such diagrams and analytical expressions for the self-energies are given in Appendix \ref{apxB}. The diagrams can be divided into four classes as it is depicted in Fig. \ref{4}. Then, the corresponding analytical expressions for the self-energies of each class are
\begin{eqnarray}
&&\Sigma^{(1)}_{(i)} = -\frac{V}{N}\sum_p \Gamma^{\mu}\cdot f_{p}\,,\label{S1}\\
&&\Sigma^{(2)}_{(ii)} = \left(\frac{V}{N}\right)^2\sum_{pq}\frac{\Gamma^{\mu}\,\Gamma^{\nu}}{2\varepsilon_p}\left(2f_{p}-1\right)\,f_q\,,\label{S2}\\
&&\Sigma^{(2)}_{(iii)} = \left(\frac{V}{N}\right)^2\left(\sum_p\Gamma^{\nu} \partial_{\omega} f |_{w = \varepsilon_p}\right)\left(\Gamma^{\mu} f_q\right)\,,\label{S3}\\
&&\Sigma^{(2)}_{(iv)}= C_{\mu\nu}\left(\frac{V}{N}\right)^2\sum_{pq}\frac{s\,s'\,s''\,\Gamma^{\mu}\,\Gamma^{\nu}}{i\omega - s'\varepsilon_p - s''\varepsilon_q - s\varepsilon_{sr}}\nonumber\\
&&\times\left[f_{s'p}\,f_{s''q}\,f_{sr} + (1-f_{s'p})(1-f_{s''q})(1 - f_{sr})\right]\,,\label{S4}
\end{eqnarray}
where $\mu,\,\nu=A,\,B,\,C$\,; $s,\,s',\,s'' = \pm 1$; $sr = sk-s'p-s''q$\,; $f_{s'p}\equiv f(s'\varepsilon_{p})$ stands for the Fermi-Dirac distribution function.
$C_{\mu\nu}$ are coefficients depending on the types of vertices in the diagram (see Appendix \ref{apxB} for details). The explicit dependencies of $\Gamma^{\mu}$ on the $k$, $p$ and $q$ are skipped but can be easily restored. It is important to note that in the above expressions, the excitation energies at the nodal points coincide with $\tilde{\xi}_p$,
\begin{eqnarray}
\label{enk_def}
\varepsilon_p = \left\{ {\begin{array}{*{20}{c}}
{\sqrt{\tilde{\xi}_p^2 + \tilde{\Delta}_p^2} > 0\,,~~~\tilde{\Delta}_p \neq 0};\\
{~~~~~~~~~~~~~~~~~\tilde{\xi}_p\,,~~~\tilde{\Delta}_p = 0,}
\end{array}} \right.
\end{eqnarray}
and can be negative. \textcolor{black} {At low temperatures we can assume $\partial_{w} f(w) |_{w = \varepsilon_p} \sim \delta(\varepsilon_p)$. The case of $\varepsilon_p=0$ corresponds to the bulk gap closing and is beyond the scope of our study. Hence, in what follows we will set $\partial_{w} f(w) |_{w = \varepsilon_p}=0$.}

The first-order normal (with the vertex index $\nu=C$) term $\Sigma^{(1)}_{(i)}$ doesn't depend on frequency and can lead to a shift of the excitation spectrum. However, due to the presence of the Fermi functions, such corrections can contribute to the self-energy at low temperatures only if $\tilde{\Delta}_p = 0$. Since the number of such points $\sim 1$, it can be seen from Eq.(\ref{S1}) that $\Sigma^{(1)}_{(i)} \sim V/N$.  

The second-order corrections $\Sigma^{(2)}_{(iv)}\sim (V\Gamma)^2$ depend on the external Matsubara frequency and contribute in both spectrum renormalization and damping. By definition given in Appendix \ref{apxB}, {$\Gamma < 1$} for majority points of the Brillouin zone (for which $\tilde{\Delta}_k \neq 0$). Hence, performing the perturbation expansion with {$V < 1$}, we have $V \Gamma < 1$.

In what follows, we will investigate chains of finite length, large enough for the formation of the Majorana zero modes, $N \sim 100$. In this case the spectrum shifts, $\delta\varepsilon^{\left(1,2\right)}$, found numerically from the terms $\Sigma^{(1)}_{(i)}$ and $\Sigma^{(2)}_{(iv)}$, respectively, can be of the same order. Comparison of such self-energy corrections shows that in the mentioned parameter window $1/N \sim V \Gamma^2$. Thus in what follows we consider both contributions.

Corrections to the self-energy of the second type $\Sigma^{(2)}_{(ii)}$ also contribute at the nodal points of the SC order parameter only. Therefore, they can be estimated as $\Sigma^{(2)}_{(ii)} \sim V^2\Gamma\,/\,N \sim \left( V \Gamma \right)^3$ that is beyond the accuracy of our consideration. Similarly, the contributions from the type iii) diagrams are also negligible due to the factor $\partial_{w} f(w) |_{w = \varepsilon_p}$. Thus, only the corrections (\ref{S1}) and (\ref{S4}), which actually have the same (second) order of smallness, will be relevant for our further consideration. 

Since the renormalized excitation spectrum, $\bar{\varepsilon}_{k}$, can be determined from the equation
\begin{equation}\label{spec_eq}
\det\left[\,\hat{g}^{-1}_{k}(i\omega)- \hat{\Sigma}_k(i\omega)\,\right]=0,
\end{equation}
it is obvious that the anomalous (off-diagonal) elements of the self-energy matrix result in the negligible fourth-order contributions. Then, to obtain $\bar{\varepsilon}_{k}$, it is sufficient to consider only normal electron-like matrix element due to the symmetry properties of $\hat{\mathcal{G}}^{(\alpha)}_{k}(\tau-\tau')$,
\begin{equation}
\left[\hat{\mathcal{G}}^{(\alpha)}_{k}(i\omega)\right]_{1,1}=\frac{1}{i\omega - \varepsilon_k - \Sigma_k(i\omega)}\,, 
\end{equation}
where $\left[\,\hat{\Sigma}_k(i\omega)\,\right]_{1,1} =  \Sigma_k(i\omega)$. According to the above arguments, the actual self-energy $\Sigma_k(i\omega)$ has the following form:
\begin{eqnarray}\label{Sigma_full}
\Sigma_{k}(\,i\omega\,) = \Sigma^{n(1)}_{k} + \sum_{j=1}^{4}\Sigma^{n(2)}_{k;\,j}(i\omega),
\end{eqnarray}
where the explicit expressions of the right-side terms are given in the Appendix\,\ref{apxB}.

\subsection{Modification of the excitation spectrum}\label{sec.3.2}

The first-order perturbation correction has only real part that is nonzero exceptionally in the nodal points with the negative excitation energy, 
\begin{multline}\label{Sigma1}
\Sigma^{n(1)}_{k}=\delta\varepsilon^{(1)}_k\\
=-2\frac{V}{N}\frac{\tilde{\xi}_k}{\varepsilon_k}\left\{\left(1-\cos k\right)f(\tilde{\xi}_{0})+\left(1+\cos k\right)f(\tilde{\xi}_{\pi})\right.\\
\left.+2\left(1+\frac{\tilde{\Delta}_1}{2\tilde{\Delta}_2}\cos k\right)f({\tilde{\xi}_{k_*}})\right\}.
\end{multline}
Such corrections lead to the fluctuation shift in the single-particle excitation energy $\delta\varepsilon^{(1)}$. The physical reason for this shift is as follows: when nodal modes with negative energies $\tilde{\xi}_0$, $\tilde{\xi}_{\pi}$ and {$\tilde{\xi}_{k_*}$} appear a quantum transition to the many-body GS occurs in the system. This added fermions renormalize the many-body spectrum due to the Coulomb interaction. Accordingly, the single-particle excitation spectrum is modified as well. This effect should be enhanced under the $V$ increase and vanish at $N \to \infty$, when the partial weights of single fermions in the many-body spectrum vanish.

Thus, the excitation spectrum is renormalized due to the coexistence of the Coulomb interaction, finite-size effects, and nodal points,
\begin{eqnarray}\label{denk1}
\varepsilon_{k} \to \bar{\varepsilon}_{k} = \varepsilon_{k} + \re\Sigma^{n\left(1\right)}_{k}. 
\end{eqnarray}
As shown in Sec.\,\ref{sec2}, quantum transitions with the filling of the nodal modes do not necessarily cause topological phase change. However, their experimental identification can be a precursor of the latter.

The spectrum $\bar{\varepsilon}_{k}$ has extrema at the points $\bar{k}=0,~\pi$ with an effective mass 
\begin{equation}\label{m_eff0}
m_{\bar{k}} = \sign(\tilde{\xi}_{\bar{k}})\left[\,\ddot{\tilde{\xi}}_{\bar{k}}+\frac{1}{\tilde{\xi}_{\bar{k}}}\left(\dot{\tilde{\Delta}}_{\bar{k}}\right)^2\,\right]^{-1}.
\end{equation}
In turn, its change caused by the many-body effects at these points has the form
\begin{align}
\delta m^{(1)}_{\bar{k}} = & -\frac{V}{N}\Big[\,
\bigl ( (1-L_{\bar{k}}) \cos\bar{k} + L_{\bar{k}} \bigr) f(\varepsilon_0) \notag \\
&- \bigl ( (1-L_{\bar{k}}) \cos\bar{k} - L_{\bar{k}} \bigr ) f(\varepsilon_{\pi}) \nonumber  \\
&- \left(\frac{\tilde{\Delta}_1}{\tilde{\Delta}_2}(1-L_{\bar{k}})\cos\bar{k} - 2L_{\bar{k}}
 \right)
f(\varepsilon_{k_*})\,\Big]^{-1},\label{m_eff}
\end{align}
where $L_{\bar{k}} =\dot{\tilde{\Delta}}_{\bar{k}}\,/ \,\tilde{\xi}_{\bar{k}}$.

It can be seen from Eq.\,(\ref{m_eff}) that the filling of each nodal mode with $k=0,\,\pi,\,k_*$ leads to a unique change in the effective mass at $\bar{k}$. Based on the expression (\ref{m_eff}), one can distinguish between different topological phases. For example, in the limit of weak superconductivity, $|\tilde{\Delta}_{1,2}| \ll |\tilde{t}_{1,2}|$, $L_{\bar{k}} \ll 1$, and $|\tilde{\Delta}_{1}|>2|\tilde{\Delta}_2|$ (i.e. $k_*$ doesn't exist), in the topologically trivial phases with the GSs $|\,\Phi_0\,\rangle$ or $|\,\Phi_3\,\rangle$ the effective mass change at $\bar{k}$ is absent $\delta m^{(1)}_{\bar{k}} \approx 0$. For the nontrivial phase with the GSs $|\,\Phi_{1,2}\,\rangle = c_{0,\pi}^+|\,\Phi_{0}\,\rangle$, the mass modification is $\delta m^{(1)}_{\bar{k}} \approx \mp (V/N)\cos \bar{k}$. 
 
According to \eqref{Sigma_full}, the spectrum changes in the second-order perturbation theory are given by the four self-energy terms. Taking into account their analytical form \eqref{Sigma2_P2} and 
making analytic continuation from $i\omega$ to $\omega+i 0^+$
it becomes obvious that for $\omega>0$ the main contribution is determined by $\Sigma_{1}^{n(2)}$. It's real and imaginary parts can be obtained using the Kramers-Kronig relations,
\begin{eqnarray}\label{Sigma_gen}
&&\im \Sigma_{1}^{n(2)}(k, \omega) = -\frac{V^2}{24\pi}\int_{-\pi}^{\pi} \dd p \sum_{i=1}^{N_q}\frac{|\Gamma^{A}_{k;p,-q_{0i},k-p+q_{0i}}|^2}{|\varepsilon'_{q_{0i}}+\varepsilon'_{k-p+q_{0i}}|}\,F_{1}, \nonumber\\
&&\re \Sigma_{1}^{n(2)}(k, \omega) = \frac{1}{\pi}{\textrm p.v.}\int_{-\infty}^{\infty} \frac{\im %\delta
\Sigma_{1}^{n(2)}(k, \omega') \dd\,\omega'}{\omega - \omega'}.
\end{eqnarray}
The quasimomenta $q_0$ are the roots of
the equation
\begin{equation}\label{q_roots}
\varepsilon_{p}+\varepsilon_{q_0}+\varepsilon_{k-p+q_0} - \omega = 0.
\end{equation}
It is seen from Eqs. (\ref{Sigma_gen})-(\ref{q_roots}) that $\im \Sigma_{1}^{n(2)}(k, \omega=0)=0$ due to the gapped bulk spectrum. 
Let us denote the denominator of the Green's function by 
\begin{eqnarray}\label{Den_G_normal_only}
D(\omega) = \omega - \varepsilon_{k} - \Sigma_k(\omega).
\end{eqnarray}
If the damping parameter is small, $\gamma \ll 1$, the following expansion is valid:
\begin{multline}
D(\omega + i\gamma) \approx D(\omega) + i \frac{\partial\,D}{\partial\,\omega}\,\gamma 
=\re D - \frac{\partial\,\im D}{\partial\,\omega}\,\gamma
\\
+i\left(\im D + \frac{\partial\,\re D}{\partial\,\omega}\,\gamma \right)=0.
\end{multline}
Then, assuming that $\left(\partial{\im D}/\partial{\omega}\right)\,\gamma \sim \gamma^2$, the expressions for the %damping 
attenuation and shift of the spectrum are as follows
\begin{eqnarray}\label{gamma_denk}
&&\gamma_{k} = \left.\frac{-\im D}{\partial{\re D}/\partial\omega}\right|_{\omega = \omega_k},~~\\
&&\delta\varepsilon_k = \omega_k - \varepsilon_k=\re \Sigma_k(\omega_k).\nonumber
\end{eqnarray}
Thus, $\delta\varepsilon_k$ and $\gamma_{k}$ are proportional to the real and imaginary parts of the self-energy at the  
frequency $\omega_k$, at which the real part of the Green's function denominator is zero, respectively. Numerical study of the dependencies of $\delta\varepsilon_k$ and $\gamma_k$ on the model parameters showed that their behavior is similar and further in this section we will discuss only $\gamma_k$. 

\begin{figure*}[htb!]\center
\includegraphics[width=1\textwidth]{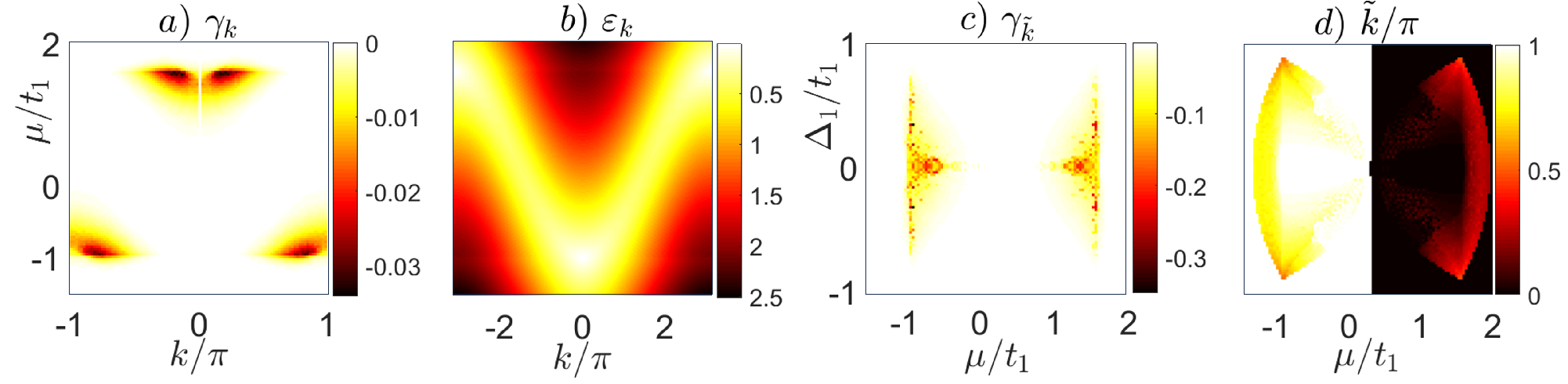}
\caption{$\left(k,~\mu\right)$ maps of the damping parameter $\gamma_k$ and mean-field spectrum $\varepsilon_k$ at $\Delta_1=0.5$ (a,b); $\left(\mu,~\Delta_1\right)$ map of the damping parameter minimum with the quasimomentum $\tilde{k}$ ($|\gamma_{\tilde{k}}|$ maximum) (c); $\left(\mu,~\Delta_1\right)$ map of the quasimomentum $\tilde{k}$ (d).  Parameters: $t_{1}=1$, $\Delta=0.626$, $t_2 = \Delta_2 = 0$, $V=0.3$, $F=G=0$, $N=100$.}\label{8}
\end{figure*}

The search for quasiparticles with finite lifetime (nonzero damping) should meet the condition of Eq.(\ref{q_roots}) with the frequency 
\begin{equation}\omega \approx \max(\varepsilon_k) > 3\min(
\varepsilon_k).
\end{equation}
Thus, the most significant renormalizations are achieved around the boundaries of topological phases, where $\min(\varepsilon_k)\ll 1$, and for quasimomenta near the maximum of spectrum. 

In Figs. \ref{8}a and %Fig.
\ref{8}b the dependencies of $\gamma_k$ and $\varepsilon_k$ on $\mu$ and $k$ are shown for the original Kitaev model, $t_{2}=\Delta_{2}$=0, at $V=0.3$ (hereinafter all energy quantities are measured in $t_1$ units). The topological transitions for such parameters are at $\mu \approx -1$ and $\mu \approx 1.5$. It can be seen that the maximum damping $|\gamma_k|$ is near $k=\pi$ at $\mu \approx -1$ as well as near $k=0$ at $\mu \approx 1.5$. Comparing Figs. \ref{8}a and 
\ref{8}b one can observe that the maxima of $|\gamma_k|$ and $\varepsilon_k$ occur at the same parametric region. 

Next, we denote the quasimomenta corresponding to the maximum attenuation as $\tilde{k}$, i.e. $|\gamma_{k=\tilde{k}}|=\max\left(|\gamma_{k}|\right)$.
As a function of $\mu$ and $\Delta_1$, in the nontrivial phase, the highest values of $|\gamma_{\tilde{k}}|$ are localized 
close to the boundary (see Fig. \ref{8}c). They penetrate deeper inside the topological region at $\Delta_{1}\ll1$. The ($\mu$, $\Delta_1$) map of $\tilde{k}$ plotted in Fig.\ref{8}d displays that below (above) half-filling these quasimomenta are mainly near $k \approx \pi$ ($k \approx 0$). 

\begin{figure}[htb!]\center
\includegraphics[width=0.5\textwidth]
{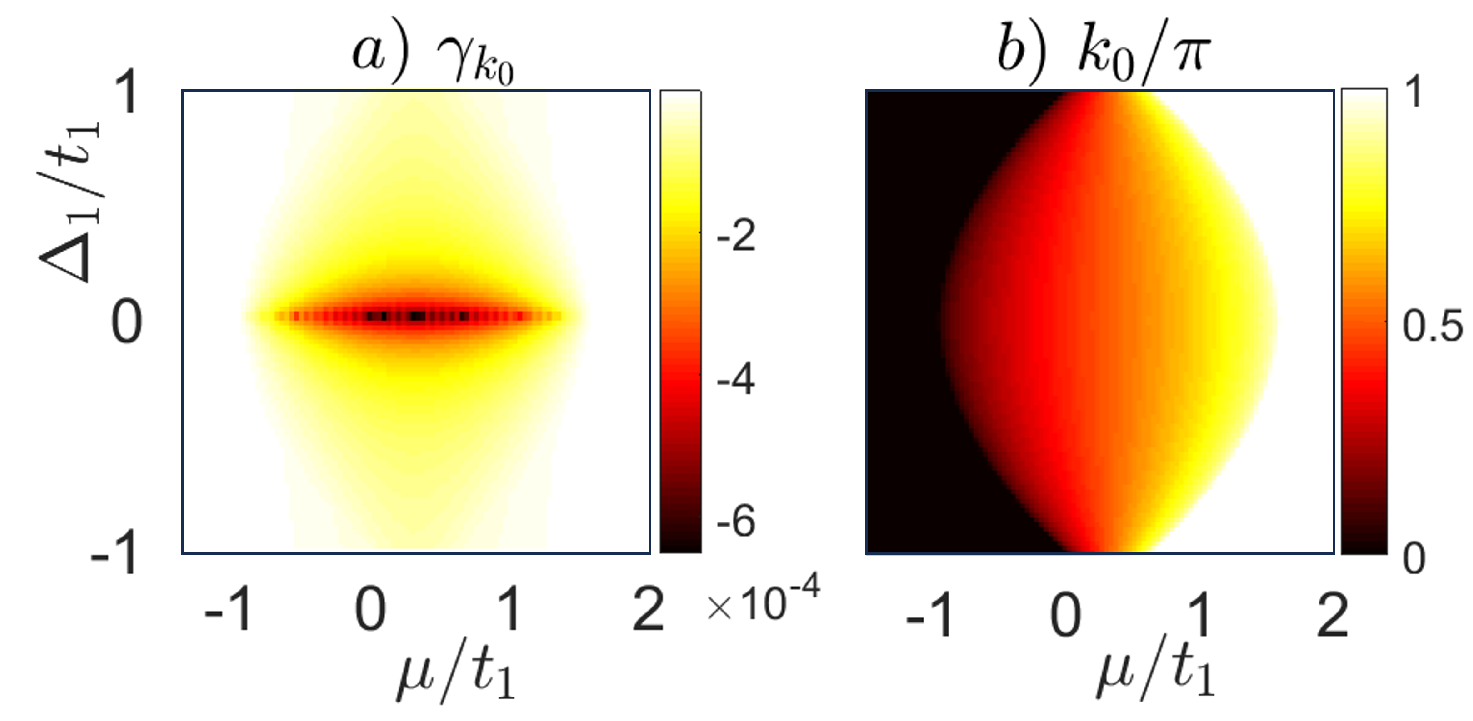}
\caption{$\left(\mu,~\Delta_1\right)$ maps of the damping parameter $\gamma_{k_0}$ at the wave vector ${k_0}$ where the mean-field spectrum has minimum  (a) and ${k_0}$ values themselves (b). Parameters are identical to the ones of Fig. \ref{8}.}\label{7}
\end{figure}

If we analyze the quasimomenta $k_0$ corresponding to the minima of the mean-field spectrum that, in the trivial phases, $k_0=0$ at $\mu<-1$ and $k_0=\pi$ at $\mu>1.5$ as it is shown at the ($\mu$, $\Delta_1$) map in Fig. \ref{7}b. In the nontrivial phase, the picture is richer. At small $\Delta_{1}$, one can observe a smooth change of ${k}_0$. The mentioned features are additionally depicted in Fig.\ref{8}b for a particular value of $\Delta_{1}$. In Fig.\ref{7}a the dependence of the damping at ${k}_0$ on $\mu$ and $\Delta_{1}$ is plotted. It is nonzero ($|\gamma_{{k}_0}|\ll|\gamma_{\tilde{k}}|$ yet, compare Figs.\ref{7}a and \ref{8}c) in the nontrivial phase and increases while the chemical potential approaches the half-filling value, $\mu=V$, and the SC pairing amplitude decreases (see dome-like pattern). In Section \ref{sec.4.2} we will show that similar conditions, i.e. $\Delta_1 \ll 1$, $\mu \approx V$, $V<1$, are favorable for the charge-fluctuation-induced topological phase. Thus, the lifetime of low-energy quasiparticles in this topological phase 
is by several orders of magnitude less than the lifetime of the corresponding excitations in the trivial phase as well as in the topological mean-field phase that is far from the fluctuation topological transition in the parametric space.

\subsection{Interaction-induced anomalous pairing}\label{sec.3.3}

While calculating the excitation spectrum, we neglected the anomalous self-energy components. However, the knowledge of $[\,\hat{\Sigma}_k(i\omega=0)\,]_{1,2}$ is necessary when calculating the topological invariant. It is useful to consider the first-order anomalous corrections to bring out some important features of such dependencies,
\begin{eqnarray}\label{Sigma_G_1}
&&\hat{\Sigma}^{(1)}_{k} = \frac{V}{N}\sum_p f_p\,\left(\,\Gamma^C_{kp;\,pk}\,\tau_z + i\,\Gamma^A_{k,p,-k;\,p}\,\tau_y\,\right)\,,\nonumber\\
&&{\delta^{(1)}\hat{\mathcal{G}}^{(\alpha)}_{k}(i\omega)} = \hat{g}_k(i\omega)\cdot{\hat{\Sigma}_{k}^{(1)}}\cdot \hat{g}_k(i\omega)\,, 
\end{eqnarray}
where analytical expressions for vertices $\Gamma^C_{kp;\,pk} \in \mathbb{R}$ and $\Gamma^A_{k,p,-k;\,p} \in i\,\mathbb{R}$ are given in Appendix \ref{apxB}.
We emphasize that in Eq. (\ref{Sigma_G_1}) the matrix $\hat{\mathcal{G}}^{(\alpha)}_{k}$ describes the perturbative corrections to the Green's function built on quasiparticle operators $\alpha_k$, see Eq.(\ref{G_alpha}). 

Summation over the Matsubara frequencies yields the sought first-order fluctuation corrections to the anomalous pairing amplitude,
\begin{gather}
\delta\Delta^{(1)}_k = T\sum_{\omega} e^{i\omega 0^+}
%\eta}
\,\left[\,{\delta^{(1)}\hat{\mathcal{G}}^{(\alpha)}_{k}(i\omega)}\,\right]^{(1)}_{1,2} 
=i\,\frac{V}{N}\,\frac{1}{\varepsilon^2_k}
\notag \\
\times
\sum_p  \left[\,{\tilde{\xi}_p}\,\tilde{\Delta}_k\,\left(1-c_{k-p}\right)-\tilde{\xi}_k\,\tilde{\Delta}_p\,c_{k-p}\,\right] \frac{f(\varepsilon_p)}{\varepsilon_p} , \label{ddelta}
\end{gather}
where $c_{k-p}=\cos(k-p)$. Note that, by analogy with the energy shift discussed in Sec.\ref{sec.3.2}, the presence of nodal points of $\tilde{\Delta}_{k}$ leads to the nonzero $\delta\Delta^{(1)}_k \sim V\,/\,N$ in the finite-size chains. 

Let us focus on the $k$-dependence of $\delta\Delta_k^{(1)}$  in the case of the Kitaev chain, for which
\begin{equation}\label{xi_Delta_Kitaev}
\tilde{\xi}_k = -\tilde{\mu} - \tilde{t}_1\,\cos k\,,~~\tilde{\Delta}_k = \tilde{\Delta}_1\,\sin k.
\end{equation} 
In this case, the antisymmetric dependence of $\delta\Delta^{(1)}_k$ %vs 
on $k$ is formed by two harmonics:
\begin{eqnarray}\label{dDelta_Kitaev}
\delta\Delta^{(1)}_k
&\sim&a_1\,\sin k + a_2\,\sin 2k, \nonumber
\end{eqnarray}
So, one can conclude that the interaction effects induce the additional harmonic in $p$-wave superconductivity. In turn, the new harmonic can provide additional nodal points of SC order parameter. As shown in Sec. \ref{sec.4.2} and Appendix \ref{apxC}, the similar nodal points, induced by charge fluctuations, occur in the second-order perturbation theory as well. 
Moreover, Sec. \ref{sec.4.2} and Appendix \ref{apxC} shows that these  nodal points lie nearby the extreme points of the mean-field spectrum,
\begin{eqnarray}
k_{0} = \arccos\left(\,\frac{-{\tilde{\mu}}\,/\,{\tilde{t}_1}}{1 - \left(\tilde{\Delta}_1\,/\,\tilde{t}_1\right)^2}\,\right):~\left.\dot{\varepsilon}\,\right|_{k=k_{0}}=0.    
\end{eqnarray}
The noted features underlie the mechanism of the fluctuation topological transitions according to the scenario similar to that discussed in Sec. \ref{sec2.2}.

\section{Topological phase transitions}\label{sec4}
\subsection{Fluctuation contribution to winding number}\label{sec.4.1}

To calculate the fluctuation contributions to the topological invariant, we use the well-known expression in terms of the Green's functions \cite{volovik-09,volovik-10,gurarie-11}
\begin{multline}
\label{N_BDI_GF}
N_{BDI}\\=\frac{1}{4\pi i}\int_{0}^{\infty}\dd \omega\int_{-\pi}^{\pi}\Tr\left[\,\tau_x\,\partial_k\hat{\mathcal{G}}^{(c)}_k\,\partial_{\omega}\left(\hat{\mathcal{G}}_k^{(c)}\right)^{-1}  \,\right.\hphantom{aaaaaa}\\
\left. -  \tau_x\,\partial_{\omega}\,\hat{\mathcal{G}}^{(c)}_k\,\partial_{k}\left(\hat{\mathcal{G}}_k^{(c)}\right)^{-1}\right] \dd k .
\end{multline}
%where $\tau_{x}$ - Pauli matrix; 
Here $\hat{\mathcal{G}}^{(c)}_k(i\omega)$ is the matrix Green's function built on the bare operators $c_k$:
\begin{multline}\label{G_c}
\hat{\mathcal{G}}^{(c)}_{k}(\tau-\tau')\\
=-\left( \begin{array}{*{20}{c}}
\langle\,T_{\tau}\,c_k(\tau)\,c_k^+(\tau')\,\rangle & \langle\,T_{\tau}\,c_k(\tau)\,c_{-k}(\tau')\,\rangle \\
\langle\,T_{\tau}\,c^+_{-k}(\tau)\,c^+_{k}(\tau')\,\rangle & \langle\,T_{\tau}\,c^+_{-k}(\tau)\,c_{-k}(\tau')\,\rangle
\end{array} \right)\\
=T\sum_{\omega} e^{-i(\tau-\tau')\omega}\,\hat{\mathcal{G}}^{(c)}_{k}(i\omega). 
\end{multline}
It can be obtained by the $SU(2)$ rotation of the matrix $\hat{\mathcal{G}}^{(\alpha)}_{k}(i\omega)$ that is valid for the arbitrary $n$-order correction,
\begin{multline}\label{Gc2Ga}
\hat{\mathcal{G}}^{(c)}_{k}=U_{mf}\,\hat{\mathcal{G}}^{(\alpha)}_{k}\,U_{mf}^+\,;~~~U_{mf} = \left(\begin{array}{*{20}{c}}
{u_k} & {-iv_k} \\
-iv_k & {u_k}
\end{array}\right).
\end{multline}

Note that, when calculating the topological invariant, we put $l_k \equiv 1$ in Eq.\,(\ref{uv_def}) and the Green's functions $\hat{\mathcal{G}}^{(c)}_{k}$ in Eq.\,(\ref{N_BDI_GF}) depend on the real frequency $i\omega \to \omega \in \mathbb{R}$.
Using its asymptotic behavior 
$$\left.\hat{\mathcal{G}}^{(c)}_k(\omega)\right|_{\omega \to \infty} \to \left(\,1\,/\,\omega\,\right)\tau_0,$$
that does not depend on the quasimomentum, the expression for the topological invariant can be rewritten as:
\begin{equation}\label{N_BDI_GF2}
N_{BDI}=\frac{1}{4\pi i}\int_{-\pi}^{\pi}\dd k\,\Tr\left.\left(\,\tau_x\,\hat{\mathcal{G}}^{(c)}_k\cdot\partial_k\left(\hat{\mathcal{G}}_k^{(c)}\right)^{-1} \,\right)\right|_{\omega = 0}.
\end{equation}
This expression involves the matrix Green's functions at zero frequency $\hat{\mathcal{G}}^{(c)}_k(\omega = 0)$. 
Let us transfer to a basis in which such a matrix is diagonal. It can be seen from Eqs. (\ref{Gc2Ga})-(\ref{N_BDI_GF2}) that for this goal it is sufficient to diagonalize the matrix
\begin{eqnarray}
\left[\hat{\mathcal{G}}^{(\alpha)}_k(\omega = 0)\right]^{-1} = \left(\begin{array}{*{20}{c}}
{X_k} & {iY_k} \\
-iY_k & {-X_k}
\end{array}\right),
\end{eqnarray}
where 
\begin{multline}\label{Xk_Yk_def}
X_k = -\varepsilon_k - \left[\,\hat{\Sigma}_k(\omega=0)\,\right]_{1,1}, ~
Y_k = i\left[\,\hat{\Sigma}_k(\omega=0)\,\right]_{1,2},
\end{multline}
and $X_k = X_{-k}$, $Y_k = -Y_{-k}$, with $X_k,\,Y_k \in \mathbb{R}$. The latter can be diagonalized by the unitary rotation
\begin{gather}
{U}_{fl}^+\,\hat{\mathcal{G}}^{(\alpha)}_{k}(\omega=0)\,{U}_{fl} = \lambda_k \tau_z\,,~~{U}_{fl} = x_k \tau_0 - iy_{k} \tau_x,\notag \\
x_k = \sqrt{\frac{1}{2}\left(1+\frac{X_k}{R_k}\right)}\,,~y_k = \sign(Y_k)\,\sqrt{\frac{1}{2}\left(1-\frac{X_k}{R_k}\right)},\notag \\
R_k = \sqrt{X_k^2 + Y_k^2}.\label{Ubar_def}
\end{gather}
Then, by combining the Eqs. (\ref{G_c})-(\ref{Ubar_def}) it is easy to get an expression for the topological invariant
\begin{align}
N_{BDI} & = \frac{1}{2\pi i}\int_{-\pi}^{\pi}\Tr\tau_x\left( U^+_{mf} \, \partial_k {U}_{mf} + %\tau_x\,
U^+_{fl}\, \partial_k {U}_{fl}\right)\dd k \notag\\
& =\frac{1}{2\pi}\oint \left(\, \dd \beta_k + \dd \bar{\beta}_k\,\right)\equiv N_{BDI}^{mf}+N_{BDI}^{fl}, 
\label{N_BDI_full}
\end{align}
where the angle $\beta_k$ is defined in Eq.(\ref{beta_def}), and the angle $\bar{\beta}_k$ is defined similarly,
\begin{equation}\label{betabar_def}
\bar{\beta}_k = \arctan\left(\frac{Y_k}{X_k}\right),~~\cos \frac{\bar{\beta}_k}{2} = x_k,~~\sin \frac{\bar{\beta}_k}{2} = y_k.
\end{equation}
Thus, contribution to the winding number due to the Coulomb fluctuations can be written by analogy with the mean-field term $N_{BDI}^{mf}$ \eqref{N_BDI_MF} as follows
\begin{equation}
\label{N_BDI_FL}
N_{BDI}^{fl} = \frac{1}{2\pi}\int_{-\pi}^{\pi}\frac{X_k \dot{Y}_k - Y_k \dot{X}_k}{X^{2}_k+Y^{2}_k}dk.
\end{equation}

\begin{figure}[htb!]\center
\includegraphics[width=0.37\textwidth]{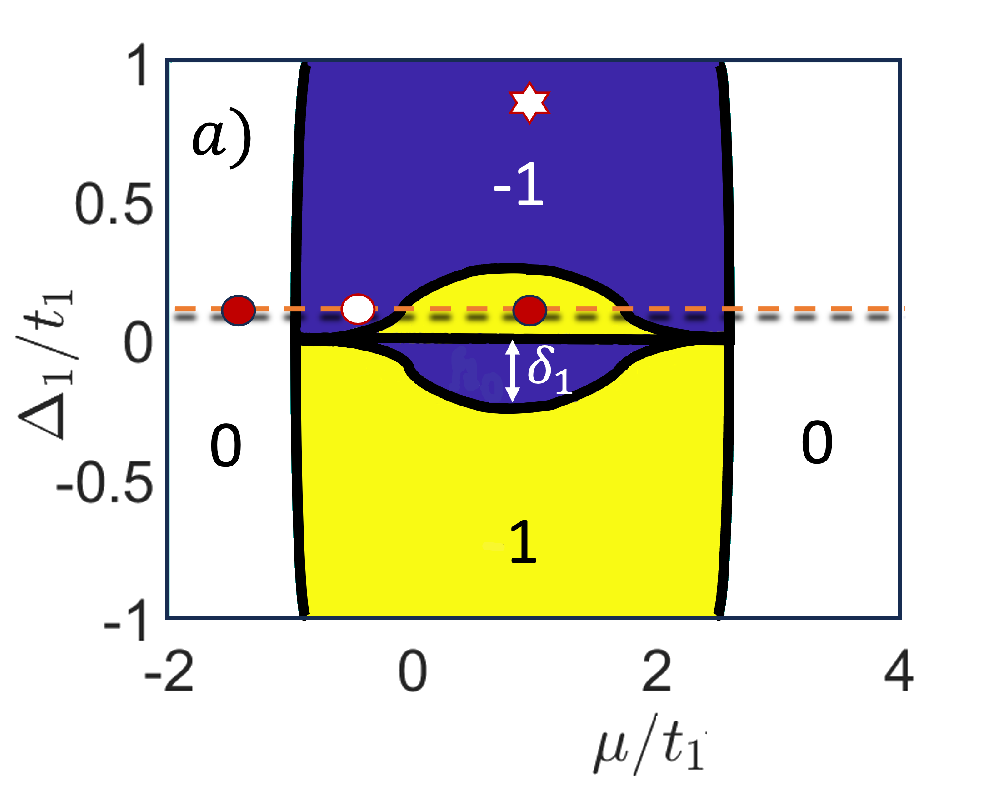}
\includegraphics[width=0.365\textwidth]{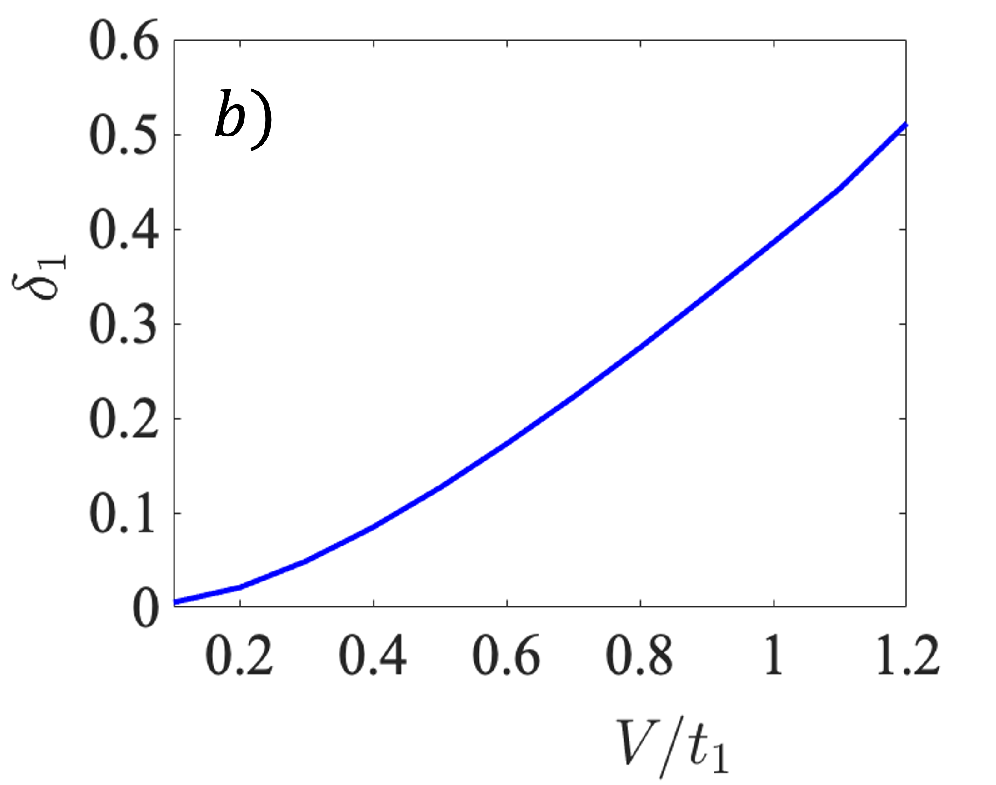}
\caption{
(Upper panel) $\left(\mu,~\Delta_1\right)$ map of the topological invariant $N_{BDI}$, Eq.\,(\ref{N_BDI_full}). 
The fluctuation contributions, Eq.\,(\ref{N_BDI_FL}) %, at 
for $V = 0.8t_1$ are taken into account. 
Pockets of height $\delta_1$, located near the line $\Delta_1 = 0$ and inside nontrivial phases, are caused precisely by such contributions. Dots, asterisk and dashed line indicate the parameters for which Fig.\,\ref{10}, Fig.\,\ref{6} and Fig.\,\ref{11} are plotted, respectively. %(b) 
(Bottom panel) $V$-dependence of the dome height $\delta_1$ which is described by the dependence $\delta_1 \approx 0.5\cdot(V\,/\,t_1)^2$.}\label{9}
\end{figure}
As it will be discussed below, such contributions can significantly modify the topological phase diagram.

\begin{figure*}[htb!]\center
\includegraphics[width=1\textwidth]{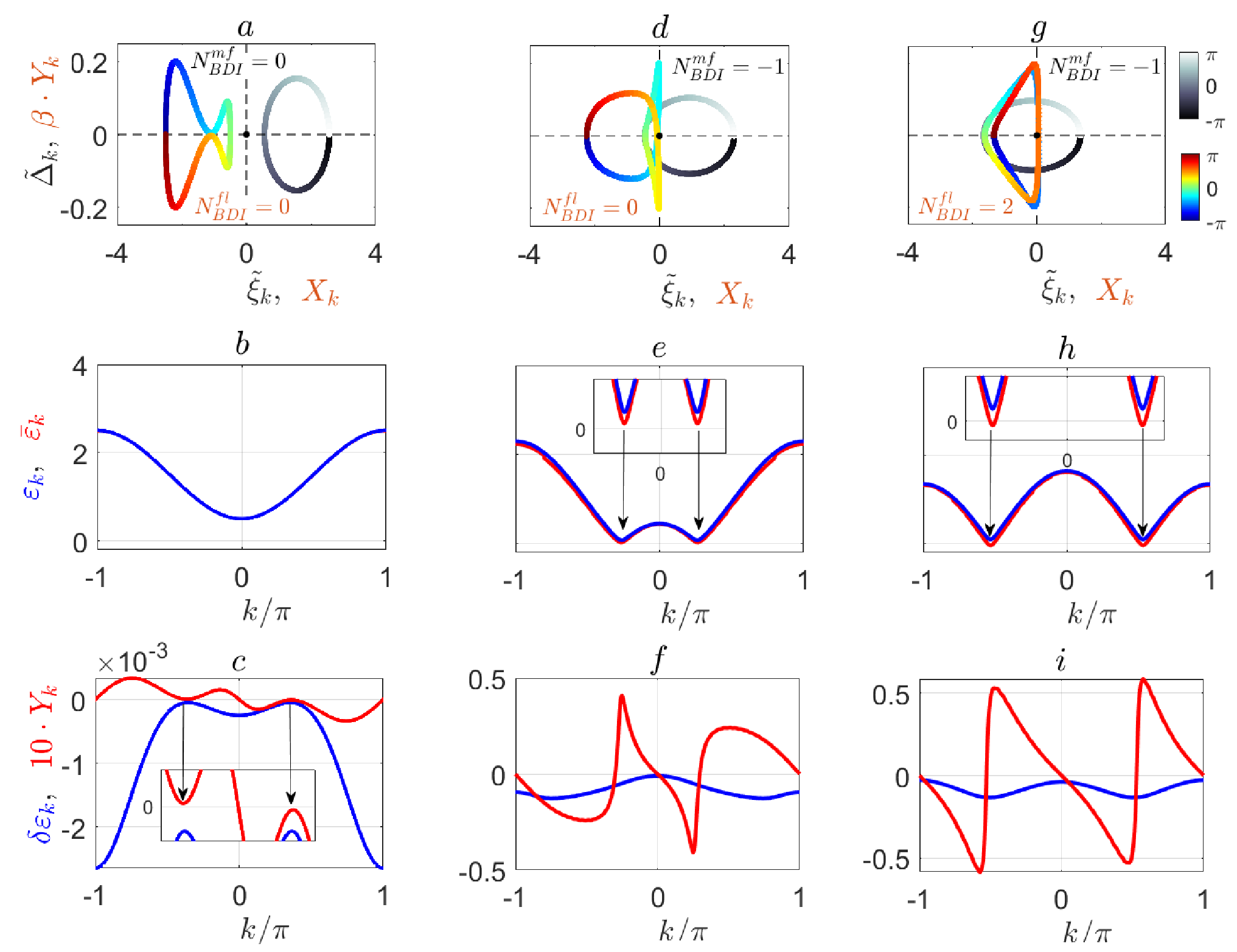}
\caption{Mechanism of topological phase transition caused by charge fluctuations. First line: $\tilde{\Delta}_k$ vs. $\tilde{\xi}_k$ (gray) and $X_k$ vs. $Y_k$ (color) dependencies forming the mean-field and fluctuation loops which winding numbers determine $N^{mf}_{BDI}$ and $N^{fl}_{BDI}$, respectively, see Eq.\,(\ref{N_BDI_full}). The points $k=-\pi$ and $k=\pi$ correspond to the black and white (blue and red) colors on the mean-field (fluctuation) loop, respectively. To make loops of the similar size, a scaling factor $\beta$ is used. Second line: quasimomentum dependencies of the mean-field, $\varepsilon_k$ (blue lines), and fluctuation-renormalized, $\bar{\varepsilon}_k = \varepsilon_k + \delta\varepsilon_k$ (red lines), quasiparticle spectra. The insets are zooms in the vicinity of the spectral minima. Third line: quasimomentum dependencies of the fluctuation-induced spectrum shift, $\delta\varepsilon_k$, as well as the scaled static part of the anomalous pairing amplitude, $Y_k$. The inset is zoom in the Brillouin zone interval where $Y_k$ is almost zero. The first, second and third columns of the plots correspond to three points in Fig.\,\ref{9}a with parameters: $t_1=1$, $V=0.8$; $\Delta_1=0.1$; $\mu = -1.5\,,\,-0.5\,,\,1$, respectively. All energy variables are in units of $t_1$.}\label{10}
\end{figure*}

\begin{figure*}[htb!]\center
\includegraphics[width=1\textwidth]{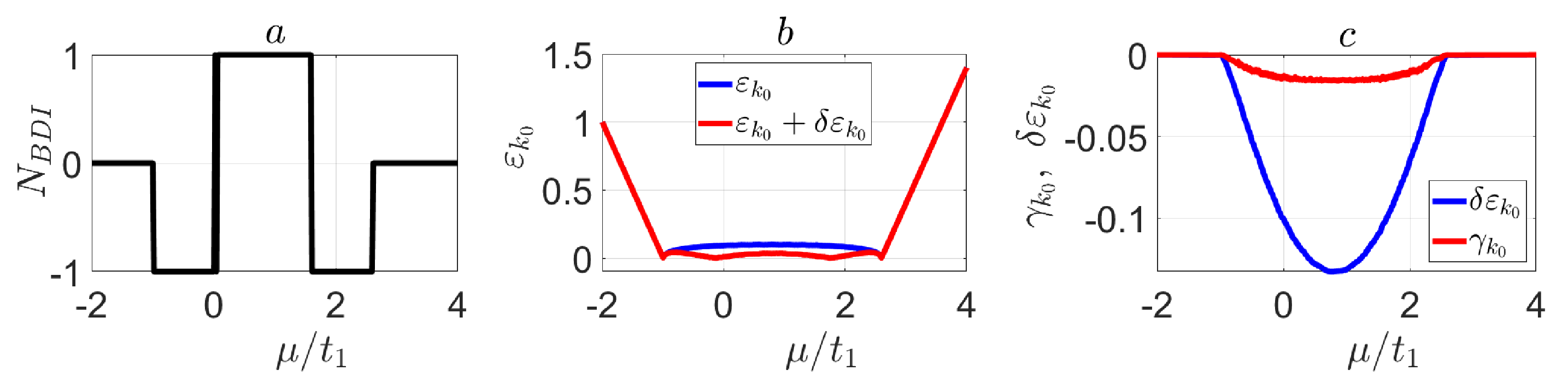}
\caption{Chemical-potential dependencies of topological invariant $N_{BDI}$, see Eq.\ref{N_BDI_full} (a); mean-field $\varepsilon_{k_0}$ and renormalized $\varepsilon_{k_0}+\delta\varepsilon_{k_0}$ spectra (b); damping parameter $\gamma_{k_0}$ and spectrum shift $\delta\varepsilon_{k_0}$ (c) at the quasimomentum $k_0$ where the mean-field spectrum has minimum. Parameters, $t_1 = 1$, $\Delta_1=0.1$, $V=0.8$, $t_{2},\Delta_2,F,G=0$, correspond to the dashed line in Fig.\,\ref{9}a. All energy variables are in units of $t_1$.}\label{11}
\end{figure*}

\subsection{Fluctuation topological phase transitions}\label{sec.4.2}

Let us consider the mechanism of the topological phase transitions conditioned by the charge fluctuations using again the standard Kitaev model as an example ($t_{2},\Delta_2,F,G=0$).
For this case, the $N_{BDI}$ map in variables $(\mu\,,\,\Delta_1)$ obtained taking into account the perturbation theory corrections up to the second order in interaction is shown in Fig.\,\ref{9}a %at 
for $V=0.8$. The mean-field term $N^{mf}_{BDI}$ in Eq. \eqref{N_BDI_full} possesses the following values:
$N^{mf}_{BDI}=-1~\textrm{if}~\{\,-1 \leq \mu \leq 2.5,~\Delta_1 > 0\,\}\,;$ $N^{mf}_{BDI}=1~\textrm{if}~\{\,-1 \leq \mu \leq 2.5,~\Delta_1 <0\,\}\,;$ $N_{BDI}=0$ else. The deviation from these values in Fig.\,\ref{9}a is due to the fluctuation contribution $N^{fl}_{BDI}$. 

It can be seen that, for example, if $0<\Delta_1 \ll 1$ and inside the mean-field nontrivial phases, the fluctuation loop with coordinates $X_{k},~Y_{k}$ can wrap the origin. Hence, they are able to give the contribution to the total winding number that is additional to the mean-field locus with coordinates $\tilde{\xi}_{k},~\Delta_{k}$. To analyze the causes and conditions of this effect, we focus on the three points on the topological diagram (circles in Fig.\,\ref{9}a) with parameters: $\Delta=0.1$\,;  $\mu = -1.5,\,-0.5,\,1$. For such points, the plots are presented in the left, central and right columns in Fig.\,\ref{10}, respectively. The graphs demonstrate: the mean-field and fluctuation loops (gray and colored curves in the first line); the mean-field and renormalized excitation spectra (blue and red curves in the second line); the spectrum shift and static part of quasiparticle anomalous pairings (blue and red curves in the third line).

When $\Delta_1=0.1$, $\mu=-1.5$ (the left point in Fig.\,\ref{9}a and Figs.\,\ref{10}a-c), the mean-field and fluctuation loops give trivial winding numbers, $N^{mf}_{BDI}=N^{fl}_{BDI}=0$. The loops intersect the lines $\tilde{\Delta}_k = 0$ and $Y_k=0$ at the symmetrical nodal points $k=0,~\pi$. The minimum distance from the loops to the origin is determined by the minima of the mean-field, $\varepsilon_k$, and renormalized, $\bar{\varepsilon}_k$, spectra, respectively, presented in Fig.\,\ref{10}b. With the parameters under consideration, such spectra are almost indistinguishable and have the minimum at $k=0$. Therefore, the distance from the mean-field loop to the origin is positive and equal to $\tilde{\xi}_0$, while the distance from the fluctuation loop to the origin is negative $X_{0} \cong -(\tilde{\xi}_0 + \delta\varepsilon_0)$. The different signs can be explained simply by the definition of $X_k$, see Eq\,.(\ref{Xk_Yk_def}). 

The shape of the mean-field loop is ellipsoidal. In turn, the fluctuation loop is more complex. This is due to the fact that the mean-field loop is described by the single $k$-harmonic (see Eq.\,(\ref{xi_Delta_Kitaev})). In opposite, the fluctuation loop is determined by the several ones. The similar effect was obtained in Sec.\,\ref{sec.3.3} for the first-order corrections to the anomalous self-energy (see Eq.\,(\ref{ddelta}) and the discussion below). Despite the oscillating behavior $Y_k$, the $Y_k$ vs $k$ dependence shows that this quantity is equal to zero only at the symmetric nodal points $k=0,~\pi$. Finally, the spectrum shift $\delta{\varepsilon}_k$ shown in Fig.\,\ref{10}c is much less than the gap and dispersion width of the mean-field spectrum (see also the inset).

Next, at the central point in Fig.\,\ref{9}a, the mean-field loop covers the origin and $N^{mf}_{BDI}=-1$ (see Fig.\,\ref{10}d). Although $N^{fl}_{BDI}=0$, the fluctuation loop contains an important new feature. In addition to the symmetric nodal points $k=0,\,\pi$, it intersects the $Y_k=0$ line at $k=\pm k^{(2)}_{*}$. The latter is clearly seen in Fig.\ref{10}f as well where the $Y_k$ vs $k$ dependence is displayed. Essentially, the appearance of $k^{(2)}_*$ is accompanied by the occurrence of new minimum points $k_0$ of the mean-field spectrum and $k^{(2)}_* \cong k_0$ (cf. Fig.\,\ref{10}f and Fig.\,\ref{10}e). The last feature was already observed for the first-order corrections in Sec.\ref{sec.3.3} and is quite general. Its justification is given in Appendix \ref{apxC}. The spectrum-shift magnitude $|\delta\varepsilon_k|$ presented in Fig.\,\ref{10}f significantly increases in comparison with the trivial phase in Fig.\,\ref{10}c. Moreover, this correction becomes comparable to the excitation energy at minimum points $\varepsilon_{k_0} = (\,\tilde{\xi}^2_{k_0} + \tilde{\Delta}^2_{k_0}\,)^{1/2}$.
As a result, the renormalized spectrum $\bar{\varepsilon}_k=\varepsilon_k + \delta\varepsilon_k$ approaches zero: $\min\left(\bar{\varepsilon}_k\right) \cong 0$, see inset in Fig.\,\ref{10}e. Note that in the presented case $\min\left(\bar{\varepsilon}_k\right) > 0$ yet. Therefore, the fluctuation loop still does not wrap the origin and $N^{fl}_{BDI} = 0$.

Finally, we consider the right point in Fig.\,\ref{9}a with $\Delta_1=0.1$, $\mu=1$ represented by the plots in Fig.\,\ref{10}g-i. Their main features are the same as in Fig.\,\ref{10}d-f with the important difference that $\min\left(\bar{\varepsilon}_k\right) < 0$. It compels the fluctuation loop to wind two times around the origin in Fig.\,\ref{10}g leading to $N_{BDI}^{fl}=2$. Thus, the total invariant  $N_{BDI}=N^{mf}_{BDI}+N^{fl}_{BDI}=1$ at $\Delta_1=0.2$ and $\mu=1$. 

Note that the occurrence of fluctuation topological transitions (FTT) accompanied by the $N^{fl}_{BDI}$ change can be effectively interpreted as the filling of the ``nodal modes" with $k=k^{(2)}_*$ and $Y_{k^{(2)}_*}=0$.  Then, we can formulate the following self-consistent mechanism of the phase transitions caused by the fluctuations: %FTT
the mean-field spectrum minima, $\varepsilon_{k_0}\ll 1$, induce nodal points $k^{(2)}_*$ of the fluctuation ``order parameter" $Y_k$ near $k_0$. In this situation, due to the Kramers-Kronig relations \eqref{Sigma_gen}, the nonzero negative shift $\delta\varepsilon_k<0$ appears. If it is enough to close and reopen the gap of the renormalized spectrum $\bar{\varepsilon}_k=\varepsilon_k + \delta\varepsilon_k$ or, in other words, to fill effective the modes with $k=k^{(2)}_*$, the fluctuating transitions occur. Mathematically, to implement the FTT, one should satisfy the following conditions:
\begin{eqnarray}\label{FTT_cond}
|\,\delta\varepsilon_{k^{(2)}_*}\,| > \varepsilon_{k^{(2)}_*}\,,~~k^{(2)}_* \cong k_0\,,~~\dot\varepsilon_{k_0} = 0. 
\end{eqnarray}

Since the behavior of the spectrum shift $\delta\varepsilon_k$ and damping parameter $\gamma_k$ is similar, the FTT emergence, in principle,  implies the finite lifetime of the low-energy excitations.  Indeed, the $\left(\,\mu\,,\,\Delta_1\,\right)$ maps of $\gamma_{k_0}$ and $N_{BDI}$ displayed in Figs. \ref{7}a  and
\ref{9}a, respectively, demonstrate qualitatively similar dome-like patterns (or pockets) near the $\Delta_1=0$ line where the FTT conditions are met. This situation is also clearly seen in Fig.\,\ref{11}c where both the $\mu$-dependencies of the damping parameter $\gamma_{k_0}$ and spectrum shift $\delta\varepsilon_{k_0}$ at the quasimomentum $k_0$ of the mean-field-spectrum minimum are provided. For the chosen $\Delta_1$, the damping and shift are nonzero throughout the nontrivial phase with $N_{BDI}\neq0$ shown in Fig.\,\ref{11}a, i.e. at $-1<\mu<2.5$. However, they rich maximum amplitudes at the center of the fluctuation-caused topological phase whose boundaries are determined by the zeros of the renormalized spectrum at $k_{0}$ (see red curve in Fig. \ref{11}b). 
However, it is worthwhile to emphasize that the presence of damping does not affect the magnitude of $N_{BDI}$ since the latter is determined by the Green's function at zero frequency. 

\begin{figure*}[htb!]\center
\includegraphics[width=1\textwidth]{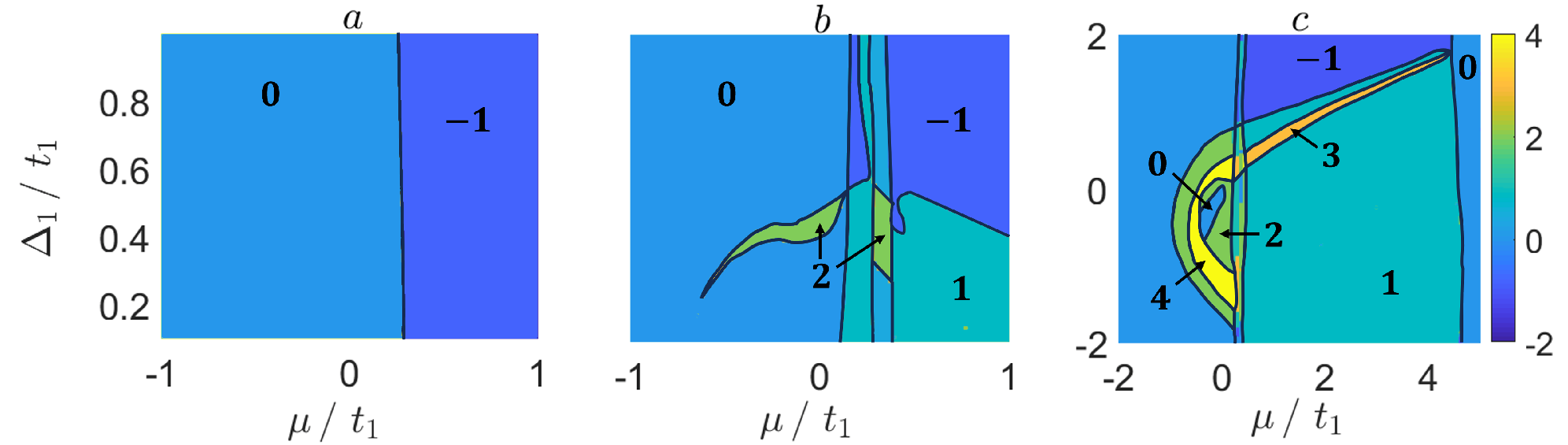}
\caption{Effect of charge fluctuations on topological phase diagram of the generalized Kitaev chain. (\,$\mu$,~$\Delta_1$\,) maps of the mean-field $N^{mf}_{BDI}$ (a) and total $N_{BDI}=N^{mf}_{BDI} +N^{fl}_{BDI}$ (b) invariants. In both cases $t_1=1$, $t_2 = -0.8$, $\Delta_2 = 0$, $V=1.2$, $\Delta_2=F=G=0$. (c) (\,$\mu$,~$\Delta_1$\,) map of  $N_{BDI}$  for the parameters of Fig.\,\ref{1}. In the latter case, the coexistence of nodal points of proximity- and fluctuation-induced SC order parameters ($k_*$ and $k_*^{(2)}$, respectively) leads to a cascade of topological transitions in a narrow window of the system parameters.}\label{12}
\end{figure*}

The maxima of $|\gamma_{k_0}|$ and $|\delta\varepsilon_{k_0}|$ correspond to the half-filling regime, $\mu=V$ and  $\tilde{\mu}=0$. Then, the value of $\Delta_1$, at which $N^{fl}_{BDI}$ changes (i.e. the dome height $\delta_1$ in the Fig.\,\ref{9}a), can be approximately found from the equation $|\,\delta\varepsilon_{k_0}\,|= \,\varepsilon_{k_0}\,$.
Numerical calculation of $\delta_1$ shows that its dependence on the interaction $V$ is $\delta_1(V)\approx V^2\,/\,2$, for $V < 1$, as expected in the developed second-order perturbation theory. Thus, at $V \ll 1$, the regions of the topological diagram modified by the fluctuation loop are very close to the line $\Delta_1=0$ (see Fig.\ref{9}b). Backed by numerical data, this feature is preserved when long-range hopping and SC pairing are taken into account: the parametric regions of fluctuation transitions lie near some mean-field topological boundaries if $V\ll t_1$ and can deviate from them significantly if $V\sim t_1$ (see Figs. \ref{9}a and \ref{12}).

Now we return to the extended Kitaev chain model. One of the consequences the fluctuation-induced nodal points give rise to is the appearance of the nontrivial phases with $|N_{BDI}|>1$ if $t_2\neq0,~\Delta_2=0$. Since the condition \eqref{k*} is not met, these phases cannot exist within the mean-field framework. However, taking into account the fluctuation effects makes their realization possible. It is easy to see this striking difference by comparing the topological phase diagrams in Figs. \ref{12}a and \ref{12}b where $N_{BDI}^{mf}$ and $N_{BDI}$ are displayed, respectively. Next, in a general case of both $t_2\neq0$ and $\Delta_2\neq0$, it is natural to expect an even richer picture of the topological phase transitions due to an interplay between the native and fluctuation-induced nonsymmetrical nodal points, $k_*$ and $k^{(2)}_*$. For example, in Fig. \ref{12}c we show the modification of the mean-field phase diagram depicted in Fig. \ref{1}. In the parametric 
range of the initially nontrivial phase where $N_{BDI}^{mf}=2$, one can observe a cascade of topological transitions. In particular, the charge fluctuations result in new phases with $N_{BDI}=4$ and $N_{BDI}=0$ since $k_*,~k^{(2)}_*$ are close to each other here.

\section{\label{sec5} Conclusions}
In this paper we have developed the theory of fluctuation topological transitions in Majorana nanowires of the BDI class with Coulomb interaction. In particular, the Kitaev chain model with long-range hopping and superconducting pairings, as well as Coulomb repulsion between neighboring fermions of the strength $V$ was considered. It was demonstrated that the topological invariant, $N_{BDI}$, built on the Green's functions, for such a system decomposes into the sum of the mean-field and fluctuation contributions: $N_{BDI} =N_{BDI}^{mf}+N_{BDI}^{fl}$. The existence of the latter is due to residual interaction between the Bogoliubov quasiparticles beyond the mean-field treatment. In the second order of perturbation theory in $V$, corrections to the Matsubara quasiparticle Green's function were calculated in the Gorkov-Nambu formalism. It turned out that the fluctuations lead to the nonzero $N_{BDI}^{fl}$ values only near the lines in parametric space where $N_{BDI}^{mf}$ changes. The maximum width of the fluctuation-modified topological diagram regions near these lines scales as $\sim V^2$.

The FTT physics is closely related to the properties of the anomalous Green's function at zero frequency -- $Y_k$. Its quasimomentum dependence is determined by several Fourier harmonics. Therefore, in addition to the symmetrical nodal points $k=0,\pi$, native to the system, the function $Y_k$ has additional ones, $k_*^{(2)}$. It is important that some of them appear near the minima of the mean-field spectrum $\varepsilon_k$. On the other hand, in such $k_*^{(2)}$ the spectral shift is negative, $\delta\varepsilon_{k_*^{(2)}}<0$. If it is sufficient to close the gap, see Eq.(\ref{FTT_cond}), then the nodal points induced by the fluctuations are effectively filled leading to the topological transitions. This mechanism is similar to what happens at the mean-field level, when the nodal points $k_*$ of the proximity-induced order parameter are filled. Moreover, if $k_*^{(2)} \approx k_*$, a cascade of topological transitions can occur with a slight change of the system parameters.
Note that an open issue in the study of the fluctuation topological phases is features of bulk-boundary correspondence. Recent works have shown that it can be violated in strongly correlated matter (see, e.g. \cite{zhao-23}). 

Considering the original Kitaev chain model \cite{kitaev-01}, we showed that the quasiparticle states with the energies near $\min(\varepsilon_k)$, which mostly contribute to the nonzero $N_{BDI}^{fl}$, have a finite lifetime, $\tau_{k}=1/\gamma_{k}$.
This time is much less than the one of the states for which $N_{BDI}^{mf} \neq 0$ and $N_{BDI}^{fl}=0$. However, it is still much longer than the lifetime of high-energy excitations near $\max(\varepsilon_k)$ which is minimal if $\Delta_{1} \ll t_{1}$ and near the borders of the mean-field nontrivial phase. 
Analysis of the mean-field-spectrum shift features $\delta\varepsilon_k$ made it possible to find the effect potentially suitable for detecting topological phase transitions in the long but finite nanowires. Since the implementation of topological phase is accompanied by the filling of “nodal modes” with quasimomenta $k=0,\pi,\pm k_*$, such “additional” fermions renormalize the excitation spectrum due to the interaction leading to the observable modification of the effective mass in the finite system, cf. Eq. (\ref{m_eff}). 

Finally, the fluctuation topological phases found in this work can be interpreted as a topological analogue of the so-called, vestigial order, implemented in quantum systems with composite order parameters \cite{fernandes-19}. In our case, the residual interactions between quasiparticles $\mathcal{H}_{int}$, cf. Eq. \eqref{H0_H1}, can be considered as corrections that describe fluctuations of the charge (the amplitude $C$) and superconducting (the amplitudes $A$ and $B$) orders. When $\varepsilon_{k_0}\ll1$, such fluctuations begin to play an important role, and ultimately lead to the destruction of topological order at $\varepsilon_{k_0}=0$. However, those fluctuations near the transition boundaries give rise to the fluctuation-induced phase that can be 
thought as a topological vestigial phase. 

To summarize, we developed the analytical theory of fluctuation topological transitions in Majorana wires of the BDI symmetry class. The key point of the theory is an idea to go beyond mean-field approximation and to treat the residual interactions of the Bogoliubov quasiparticles. The residual interactions were considered as perturbations when calculating the diagram series for quasiparticle Green's functions. This made it possible to explicitly extract the contribution to the topological invariant due to charge correlations and to calculate a topological phase diagram taking the latter into account. The effective filling of the nodal points in the superconducting order parameter
induced by the residual interactions was proposed as a mechanism of the fluctuation topological transitions. In addition, we studied the interaction effects on the quasiparticle spectrum.  

A further extension of the work would be to study a disorder effect on the fluctuation topological transitions. Taking into account the results of recent studies \cite{mcginley-17,kells-18,karcher-19,francica-23}, one can expect even richer picture of the topological transitions in the model (\ref{Ham_Kit_gen}) due to, e.g. interplay of disorder and nonlocal interactions $\sim F,\,G$.

\begin{acknowledgments}
We thank D.~M.~Dzebisashvili, A.~O.~Zlotnikov, and, especially, Yu.~G.~Makhlin, for fruitful discussions. S.V.A. is grateful to Landau Institute for Theoretical Physics for hospitality. The reported study was supported by Russian Science Foundation, project No. 22-42-04416.
\end{acknowledgments}
  
\appendix

\section{\label{apxA} Derivation of the effective model}

Let us rewrite the original Hamiltonian $\mathcal{H}_W$ as a sum of terms of zero, first and second order of smallness, 
\begin{eqnarray}\label{H_tot}
\mathcal{H}_W = \mathcal{H}_{0} + \mathcal{V}_1 + \mathcal{V}_2.
\end{eqnarray}
Here $\mathcal{H}_{0}$ is an unperturbed Hamiltonian and $\mathcal{V}_j$ $(j=1,2)$ -- operators describing weak interactions.

It is convenient to choose many-body eigenstates $|\,m\,\rangle$ of the Hamiltonian $\mathcal{H}_{0}$ as a basis in the Hilbert space of the operator $\mathcal{H}$: $\mathcal{H}_{0}|\,m\,\rangle=E_m|\,m\,\rangle$. 
An essential assumption allowing us to develop the perturbation theory is an existence of a large energy gap in the $E_m$ spectrum \cite{bir-74}.
The subspace below the gap (the, so called, ``low-energy"\ sector of the Hilbert space) is denoted as $\mathcal{M}$. The corresponding eigenstates and eigenvalues are
enumerated by $m$. 
To count the eigenstates above the gap in the ``high-energy"\ sector $\mathcal{L}$, the symbol $l$ is used.
Note that both $|m\rangle \in \mathcal{M}$ and $|l\rangle \in \mathcal{L}$ can be degenerate but not all the eigenvalues $E_m$ (and as well $E_l$) must necessarily be equal to each other.

Using the many-body states $|m\rangle$ we can define projection operator $P$ onto the low-energy sector $\mathcal{M}$ as: 
$P = \sum_{m \in \mathcal{M}} X^{mm}$,  
with $X^{mm}=|m\rangle\langle m|$ being the Hubbard operators. 
The projection operator allows to divide interactions $\mathcal{V}_j$ ($j=1,2$) in the Hamiltonian (\ref{H_tot}) into two parts: $\mathcal{V}_j=\mathcal{\bar{V}}_j+\mathcal{\bar{\bar{V}}}_j$. The first part ${\mathcal{\bar{V}}_j}$ consisting of two terms,
\begin{eqnarray}\label{H1}
\mathcal{\bar{V}}_j = P\,\mathcal{V}_j\,P+(1-P)\,\mathcal{V}_j\,(1-P),
\nonumber\\
P\,\mathcal{V}_j\,P = \sum_{m,m'\in \mathcal{M}}\left(\mathcal{V}_j\right)_{m,m'}\hphantom{l}X^{mm'},\nonumber\\
(1-P)\,\mathcal{V}_j\,(1-P)=\sum_{l,l'\in \mathcal{L}}\left(\mathcal{V}_j\right)_{l,l'}\hphantom{l}X^{ll'},
\end{eqnarray}
does not mix the low- and high-energy sectors of the Hilbert space and, hence, is called a diagonal part. 
The second part $\mathcal{\bar{\bar{V}}}_j$ also consisting of two terms
\begin{eqnarray}\label{H2}
\mathcal{\bar{\bar{V}}}_j = (1-P)\,\mathcal{V}_j\,P+P\,\mathcal{V}_j\,(1-P),
\nonumber\\
(1-P)\,\mathcal{V}_j\,P = \sum_{\substack{m \in \mathcal{M} \\ l \in \mathcal{L}}}
\left(\mathcal{V}_j\right)_{l,m}\hphantom{l}X^{lm},
\nonumber\\
P\,\mathcal{V}_j\,(1-P) = \sum_{\substack{m \in \mathcal{M} \\ l \in \mathcal{L}}}
\left(\mathcal{V}_j\right)_{m,l}\hphantom{l}X^{ml},
\end{eqnarray}
is nondiagonal because of mixing the sectors $\mathcal{M}$ and $\mathcal{L}$. In Eqs. (\ref{H1}) and (\ref{H2}) the matrix elements $\langle m|\mathcal{V}_j |l \rangle$ are denoted as $\left(\mathcal{V}_j\right)_{m,l}$.

Consider the following unitary transformation of the Hamiltonian $\mathcal{H}$:
\begin{eqnarray}\label{U_trans}
\mathcal{H}\to \tilde{\mathcal{H}}=e^{-S}\,\mathcal{H}\,e^{S}&=&\mathcal{H} + [\,\mathcal{H}, S\,] + \frac{1}{2}[\,[\,\mathcal{H},\,S\,],\,S\,] \nonumber\\ &+&\frac{1}{6}[\,[\,[\,\mathcal{H},\,S\,],\,S\,],S\,]+\ldots
\end{eqnarray}
We will assume that the operator $S$ in (\ref{U_trans}) is nondiagonal 
\begin{eqnarray}\label{S_tot_X}
S &=& \sum_{\substack{m \in \mathcal{M} \\ l \in \mathcal{L}}}
\left[\left(S\right)_{m,l}\hphantom{l}X^{ml} + \left(S\right)_{l,m}\hphantom{l}X^{lm} \right],
\end{eqnarray}
and its decomposition starts with terms of the first order of smallness:
\begin{eqnarray}\label{S_tot}
S &=& S_1+S_2+S_3 + \ldots.
\end{eqnarray}
 
Substituting the expressions (\ref{H_tot}) and (\ref{S_tot}) into the series (\ref{U_trans}), we retain only the terms whose order of smallness is not higher than three.
Next, to get rid of the nondiagonal terms in $\tilde{\mathcal{H}}$, the following conditions on the operators $S_1$ and $S_2$ have to be imposed:
\begin{eqnarray}\label{S1_eq}
\mathcal{\bar{\bar{V}}}_1 + [\, \mathcal{H}_{0}\, , \, S_1\,] = 0,
\\ \label{S2_eq}
\mathcal{\bar{\bar{V}}}_2 + [\,\mathcal{H}_{0}\,,\, S_2\,] + [\, \mathcal{\bar{V}}_1   ,\, S_1\,]= 0.
\end{eqnarray}
Then, the matrix elements of $S_1$ can be found from the operator equation (\ref{S1_eq}),
\begin{eqnarray}\label{S1_eq_matr}
\left(S_1\right)_{m,l} = \frac{\left(\mathcal{V}_1\right)_{m,l}}{E_{l} - E_{m}}\,,
~
\left(S_1\right)_{l,m} = -\frac{\left(\mathcal{V}_1\right)_{l,m}}{E_{l} - E_{m}}\,.
\end{eqnarray}
Here we took advantage of the equalities
$\mathcal{H}_{0}|\, m \, \rangle = E_{m}|\, m \, \rangle$ and
$\mathcal{H}_{0}|\, l \, \rangle = E_{l}|\, l \, \rangle$. 

Projecting out the high-energy processes the operators acting within the low-energy sector $\mathcal{M}$ of the Hilbert space are only left. Thus, the general form of the required effective Hamiltonian is
\begin{eqnarray}\label{H_eff}
\mathcal{H}_{eff} =P\,\mathcal{H}\,P ~~~~~~~~~~~  \nonumber\\ 
 +\frac{1}{2}\,P\left( \big[ \,\mathcal{\bar{\bar{V}}}_1\, , \, S_1 + S_2\, \big] +  \big[ \,\mathcal{\bar{\bar{V}}}_2\, , \, S_1 \big] \right)P.
\end{eqnarray}
If we consider the spin-polarized SW the projection operator on the sector $\mathcal{M}$ can be written in the form:
\begin{eqnarray}\label{P}
P = \prod_{f}\left(\,|\,f\,,0\,\rangle\langle \,f,0\,|+|\,f,\,\uparrow\,\rangle\langle\, f,\,\uparrow\,|\,\right).
\end{eqnarray}

The performed calculations for the original model (\ref{Ham_wire}) allow us to find the effective Hamiltonian (\ref{Ham_Kit_gen}) with the following parameters of effective interactions:
\begin{eqnarray}
\label{Param}
&~&\frac{E_{0}}{N}=\frac{\Delta_{1s}^{2}}{2\mu}-\frac{\Delta_s^2}{U-2\mu}, \\
&~&\mu = \mu + h + \frac{\alpha^2}{4h} + \frac{2\Delta_{1s}^2}{\mu} + \frac{2\Delta_{1s}^{2}-\Delta^2}{U-2\mu},\nonumber\\
&~&t = t_0 -  \frac{4\Delta_s\Delta_{1s}}{U-2\mu};~t_{1}= \frac{\Delta^2_{1s}}{\mu}-\frac{\alpha^2}{4h},\nonumber\\
&~&\Delta=\frac{\alpha\Delta_s}{U-2\mu}+\frac{\alpha\Delta_s}{U+2h};~~\Delta_{1}=\frac{\alpha\Delta_{1s}}{2}\left(\frac{1}{h}-\frac{1}{\mu}\right),\nonumber
\end{eqnarray}
where $E_0$ gives corrections to the ground state energy due to the virtual creations and annihilations of Cooper pairs. This term is omitted in (\ref{Ham_Kit_gen}). Taking into account the Hubbard repulsion $ U $ leads to the effective interactions between spinless fermions. On the one hand, a repulsive interaction between fermions with amplitude $V$ is induced. On the other hand, nonzero $U$ leads to three-centered charge-correlated hoppings and SC pairings of fermions belonging to the secondary coordination sphere (parameters $ F $ and $ G $, respectively). The dependence of such parameters on the SW parameters reads 
\begin{eqnarray}
\label{V_eff}
&~&{V}= V_0 {+} \frac{(\alpha/2)^2}{2h}-\frac{(\alpha/2)^2}{U{+}2h}+\frac{\Delta_{1s}^2}{2\mu}+\frac{\Delta_{1s}^2}{U{-}2\mu},\\
\label{F_eff}
&~&F=\frac{(\alpha/2)^2}{U+2h}-\frac{(\alpha/2)^2}{2h}+\frac{\Delta_{1s}^{2}}{2\mu}+\frac{\Delta_{1s}^{2}}{U-2\mu},\\
\label{G_eff}
&~&G = \frac{\alpha\Delta_{1s}}{2}\Bigg(\frac{1}{2h} - \frac{1}{U{+}2h} - \frac{1}{2\mu} - \frac{1}{U{-}2\mu}  \Bigg).
\end{eqnarray}

{Note, that the three-center interactions can be considered as the operators of the number of high-energy states deformed when projected onto the low-energy sector
\begin{eqnarray}\label{H3_deform}
\mathcal{H}_3 = P\,e^{-S}\,\left(\,X^{\downarrow\downarrow} + X^{22}\,\right)\,e^{S}\,P.\end{eqnarray}}

\section{\label{apxB} Analytic expressions for vertices and self-energy contributions}

Symmetrization of the interaction amplitudes with respect to the quasimomenta results in the following vertices $\Gamma^{A,B,C}$: 
\begin{eqnarray}
&&\Gamma^{(A)}_{k;p,q,k-p-q}\notag \\
&&=2i\,(u_k\,v_p\,u_{q}\,u_{k-p-q}+v_k\,u_p\,v_{q}\,v_{k-p-q})\left(c_{k-q}-c_{p+q}\right)\nonumber\\ 
&&\,+\,2i\,(u_k\,u_p\,v_{q}\,u_{k-p-q}+v_k\,v_p\,u_{q}\,v_{k-p-q})\left(c_{p+q}-c_{k-p}\right)\nonumber\\
&&\,+\,2i\,(u_k\,u_p\,u_{q}\,v_{k-p-q}+v_k\,v_p\,v_{q}\,u_{k-p-q})\left(c_{k-p}-c_{k-q}\right),\notag \\
&&\label{gamma_A1}
\end{eqnarray}
\begin{eqnarray}
&&\Gamma^{(A)}_{k+p+q;q,p,k}\notag\\
&&=2i\,(u_k\,u_p\,v_{q}\,u_{k+p+q}+u_q\,v_p\,v_{k}\,v_{k+p+q})\left(c_{k+q}-c_{p+q}\right)\nonumber\\ 
&&\,+\,2i\,(u_k\,u_q\,v_{p}\,u_{k+p+q}+u_p\,v_k\,v_{q}\,v_{k+p+q})\left(c_{p+q}-c_{k+p}\right)\nonumber\\
&&\,+\,2i\,(u_p\,u_q\,v_{k}\,u_{k+p+q}+u_k\,v_q\,v_{p}\,v_{k+p+q})\left(c_{k+p}-c_{k+q}\right),\nonumber
\\
&& \label{gamma_A4}
\end{eqnarray}
\begin{eqnarray}
&&\Gamma^{(C)}_{k,p;q,k+p-q}=\Gamma^{(C)}_{k+p-q,q;p,k}\notag \\
&&=2\,(u_k\,u_p\,u_{q}\,u_{k+p-q}+v_k\,v_p\,v_{q}\,v_{k+p-q})\left(c_{p-q}-c_{k-q}\right)\nonumber\\ 
&&\,+\,2\,(u_k\,u_{k+p-q}\,v_p\,v_{q}+u_p\,u_q\,v_{k}\,v_{k+p-q})\left(c_{k+p}-c_{p-q}\right)\nonumber\\
&&\,+\,2\,(u_p\,u_{k+p-q}\,v_k\,v_{q}+u_k\,u_q\,v_{p}\,v_{k+p-q})\left(c_{k-q}-c_{k+p}\right),\nonumber\\
\label{gamma_C1}
\end{eqnarray}
\begin{eqnarray}
&&\Gamma^{(B)}_{k,p,q,-k-p-q}\notag \\
&&=2\,(u_k\,u_q\,v_{p}\,v_{-k-p-q}+u_p\,u_{-k-p-q}\,v_{k}\,v_{q})\left(c_{p+q}-c_{k+p}\right)\nonumber\\ 
&&\,+\,2\,(u_k\,u_{p}\,v_q\,v_{-k-p-q}+u_q\,u_{-k-p-q}\,v_{k}\,v_{p})\left(c_{k+q}-c_{p+q}\right)\nonumber\\
&&\,+\,2\,(u_k\,u_{-k-p-q}\,v_p\,v_{q}+u_p\,u_q\,v_{k}\,v_{-k-p-q})\left(c_{k+p}-c_{k+q}\right),\nonumber\\
&& \label{gamma_B}
\end{eqnarray}
\begin{figure}[htb!]\center
\includegraphics[width=0.48\textwidth]{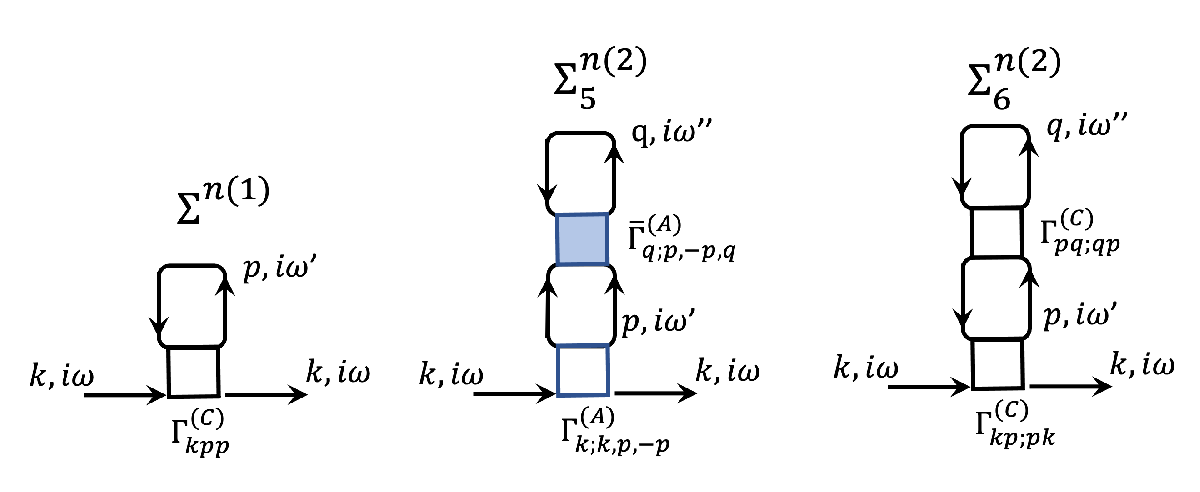}
\caption{Irreducible diagrams containing fermionic loop and appearing in the first and second orders of the perturbation theory for the normal Green's function.}\label{15}
\end{figure}

Analysis of the diagrammatic series for the normal electron-like Green's function up to the second order results in three diagrams containing fermionic loops (see Fig. \ref{15}) and leads to the following self-energy terms: 
\begin{eqnarray}\label{Sigma2_P1}
&&\Sigma^{n(1)}=-\frac{V}{N}\sum_{p}\,\Gamma^{(C)}_{k,p;p,k}\,f_p,\\
&&\Sigma^{n(2)}_{5}=\left(\frac{V}{N}\right)^2\sum_{pq}\frac{\Gamma^{(A)}_{k;k,p,-p}\,\overline{\Gamma}^{(A)}_{q;p,q,-p}}{2\varepsilon_p}\left(1-2f_p\right)\,f_q,\nonumber\\
&&\Sigma^{n(2)}_{6}=\left(\frac{V}{N}\right)^2\Bigl(\sum_{p}\Gamma^{(C)}_{k,p;p,k}\partial_{\omega} f |_{w = \varepsilon_p}\Bigr)\Bigl(\,\sum_{q}\Gamma^{(C)}_{p,q;q,p}\,f_q\Bigr)\,,\nonumber
\end{eqnarray}
where $f_{p,q}\equiv f(\varepsilon_{p,q})$ - Fermi-Dirac distribution functions. At low temperatures and in the thermodynamic limit, such corrections are exponentially suppressed due to the Fermi functions. For finite $N$, they can give contributions of the order $\sim V/N$, $\sim V^2/N$ and $\sim (V/N)^2$.

\begin{figure}[t]\center
\includegraphics[width=0.49\textwidth]{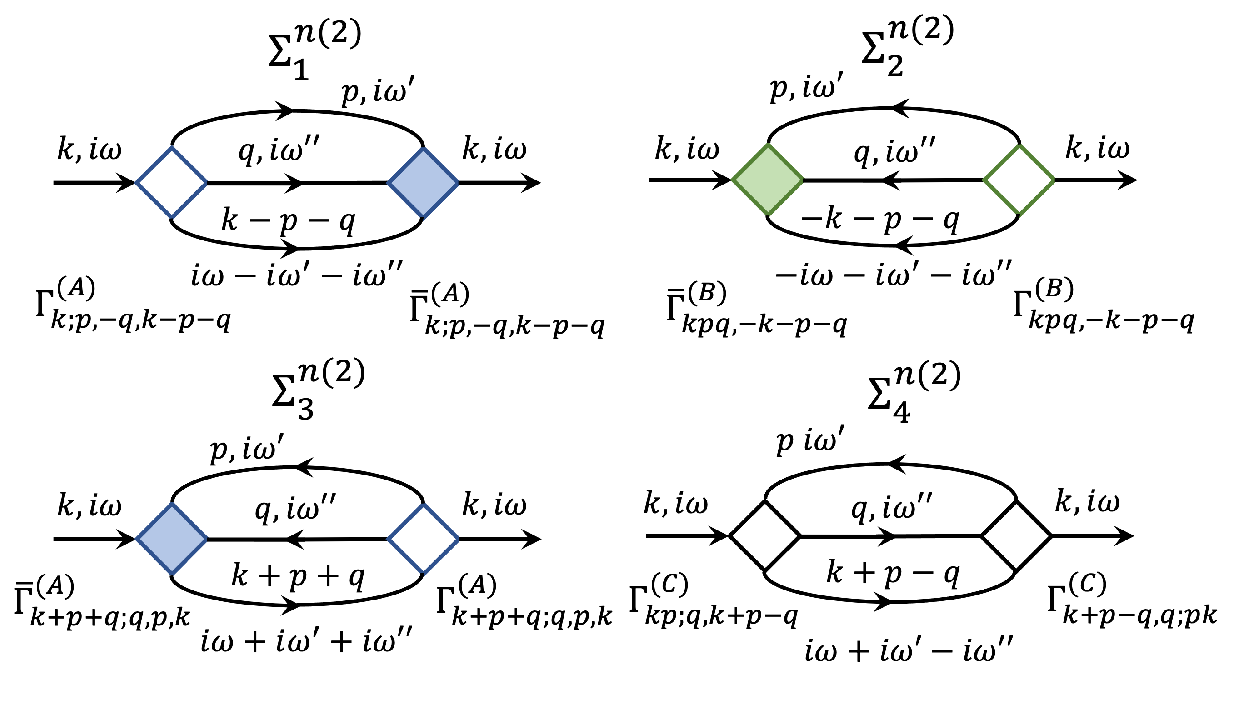}
\caption{Irreducible diagrams leading to the second-order corrections to the normal self-energy.}\label{16}
\end{figure}
The rest of diagrams for the normal Green's function is loopless as it can be seen in Fig. \ref{16}. The contributions from them to the normal self-energy are given by
\begin{eqnarray}\label{Sigma2_P2}
&&\Sigma_{1}^{n(2)} =\frac{1}{6}\left(\frac{V}{N}\right)^2\sum_{p,q} \frac{\left|\Gamma^{(A)}_{k;p,q,k-p-q}\right|^{2}F_{1}}{i\omega-\varepsilon_p - \varepsilon_q - \varepsilon_{k-p-q}}\,,\nonumber\\
&&\Sigma_{2}^{n(2)} = -\frac{1}{6}\left(\frac{V}{N}\right)^2\sum_{p,q} \frac{\left|\Gamma^{(B)}_{k,p,q,-k-p-q}\right|^{2}F_{2}}{i\omega+\varepsilon_p + \varepsilon_q + \varepsilon_{-k-p-q}}\,,\nonumber\\
&&\Sigma_{3}^{n(2)} =\frac{1}{2}\left(\frac{V}{N}\right)^2\sum_{p,q} \frac{\left|\Gamma^{(A)}_{k+p+q;q,p,k}\right|^{2}F_{3}}{i\omega+\varepsilon_p + \varepsilon_q - \varepsilon_{k+p+q}}\,,\nonumber\\
&&\Sigma_{4}^{n(2)} = \frac{1}{2}\left(\frac{V}{N}\right)^2\sum_{p,q} \frac{\left|\Gamma^{(C)}_{k,p;q,k+p-q}\right|^{2}F_{4}}{i\omega+\varepsilon_p - \varepsilon_q - \varepsilon_{k+p-q}}\,,\nonumber\\
\end{eqnarray}
where combinations of the Fermi functions are
\begin{eqnarray}
&&F_{1}=f_{p}\,f_{q}\,f_{k-p-q}+\left[1-f_{p}\right]\,\left[1-f_{q}\right]\,\left[1-f_{k-p-q}\right],\notag\\
&&F_{2}=f_{p}\,f_{q}\,f_{-k-p-q}+\left[1-f_{p}\right]\,\left[1-f_{q}\right]\,\left[1-f_{-k-p-q}\right],\nonumber\\
&&F_{3}=f_{p}\,f_{q}\,\left[1-f_{k+p+q}\right]+\left[1-f_{p}\right]\,\left[1-f_{q}\right]\,f_{k+p+q},\nonumber\\
&&F_{4}=\left[1-f_{p}\right]f_{q}f_{k{+}p{-}q}+f_{p}\,\left[1-f_{q}\right]\,\left[1-f_{k{+}p{-}q}\right].\label{F_ABC}
\end{eqnarray}
These corrections determine both the excitation spectrum shift and quasiparticle damping. 

\begin{figure}[t]\center
\includegraphics[width=0.5\textwidth]{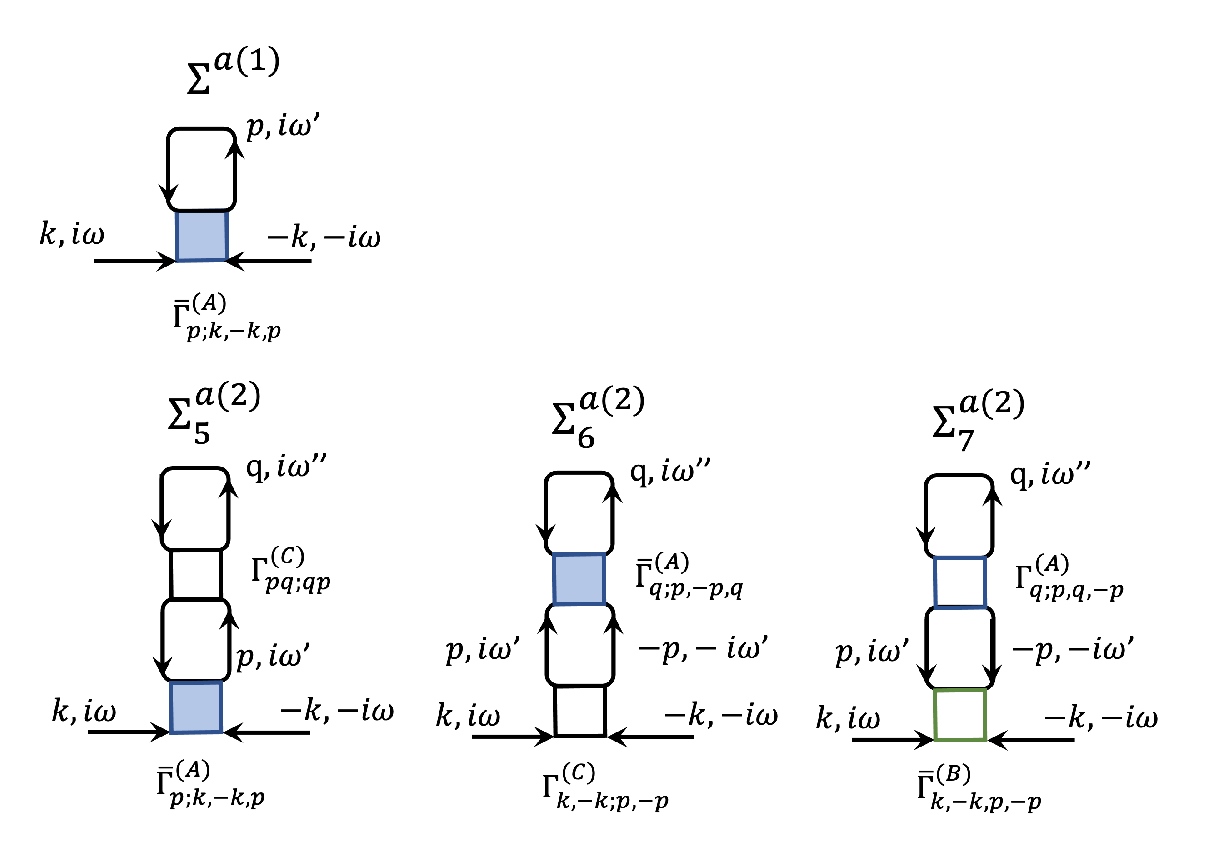}
\caption{Diagrams with fermionic loop occurring in the first and second orders of the perturbation
theory for the anomalous Green’s function.}\label{S12anom_loop}
\end{figure}

\begin{figure}[t]\center
\includegraphics[width=0.49\textwidth]{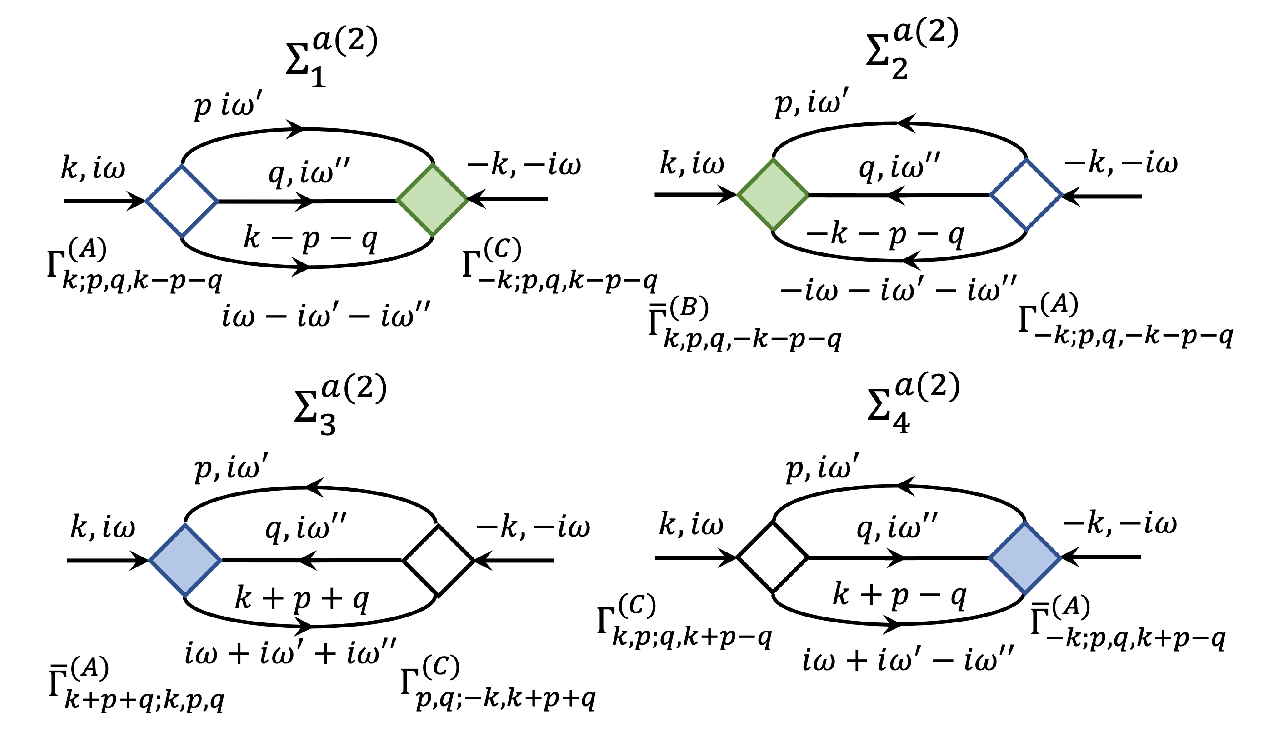}
\caption{Irreducible diagrams leading to the second-order corrections to the anomalous self-energy.}\label{S12anom_loopless}
\end{figure}

By analogy, the corrections to the anomalous self-energy from the loop-containing diagrams displayed in Fig. \ref{S12anom_loop} are
\begin{eqnarray}\label{Sigma12anom_l}
&&\Sigma^{a(1)}=-\frac{V}{N}\sum_{p}\,\Gamma^{(A)}_{k,p,-k;p}\,f_p,\\
&&\Sigma_{5}^{a(2)}=\left(\frac{V}{N}\right)^2\sum_{pq}\frac{\Gamma^{(C)}_{k,-k;p,-p}\,\Gamma^{(A)}_{p,q,-p;q}}{2\varepsilon_p}\left(1-2f_p\right)\,f_q,\nonumber\\
&&\Sigma_{6}^{a(2)}=\left(\frac{V}{N}\right)^2\Bigl(\sum_{p}\Gamma^{(A)}_{k,p,-k;p}\partial_{\omega} f |_{w = \varepsilon_p}\Bigr)\Bigl(\,\sum_{q}\Gamma^{(C)}_{p,q;q,p}\,f_q\Bigr)\,,\nonumber\\
&&\Sigma_{7}^{a(2)}=\frac{1}{2}\left(\frac{V}{N}\right)^2\sum_{pq}\frac{\Gamma^{(B)}_{k,p,-p,-k}\,\Gamma^{(A)}_{p,q,-p;q}}{2\varepsilon_p}\left(1-2f_p\right)\,f_q.\nonumber
\end{eqnarray}

The corresponding contributions from the second-order loopless diagrams shown in Fig. \ref{S12anom_loopless} read
\begin{eqnarray}\label{Sigma12anom_ll}
&&\Sigma_{1}^{a(2)}=\frac{1}{6}\left(\frac{V}{N}\right)^2\sum_{p,q} \frac{\Gamma^{(B)}_{k-p-q,p,q,-k}\,\Gamma^{(A)}_{p,q,k-p-q;k}}{i\omega-\varepsilon_p - \varepsilon_q - \varepsilon_{k-p-q}}\,F_{1},\nonumber\\
&&\Sigma_{2}^{a(2)} =-\frac{1}{6}\left(\frac{V}{N}\right)^2\sum_{p,q} \frac{\Gamma^{(B)}_{k,p,q,-k-p-q}\,\Gamma^{(A)}_{p,q,-k-p-q;-k}}{i\omega+\varepsilon_p + \varepsilon_q + \varepsilon_{-k-p-q}}\,F_{2},\nonumber\\
&&\Sigma_{3}^{a(2)} =\frac{1}{2}\left(\frac{V}{N}\right)^2\sum_{p,q} \frac{\Gamma^{(A)}_{k,p,q;k+p+q}\Gamma^{(C)}_{k+p+q,-k;p,q}}{i\omega+\varepsilon_p + \varepsilon_q - \varepsilon_{k+p+q}}\,F_{3},\nonumber\\
&&\Sigma_{4}^{a(2)} = -\frac{1}{2}\left(\frac{V}{N}\right)^2\sum_{p,q} \frac{\Gamma^{(A)}_{-k,q,k+p-q;p}\Gamma^{(C)}_{k,p;q,k+p-q}}{i\omega+\varepsilon_p - \varepsilon_q - \varepsilon_{k+p-q}}\,F_{4}.\nonumber\\
\end{eqnarray}

\section{\label{apxC} Vicinity of mean-field spectrum minima and fluctuation-induced nodal points}

\begin{figure}[htb!]\center
\includegraphics[width=0.34\textwidth]{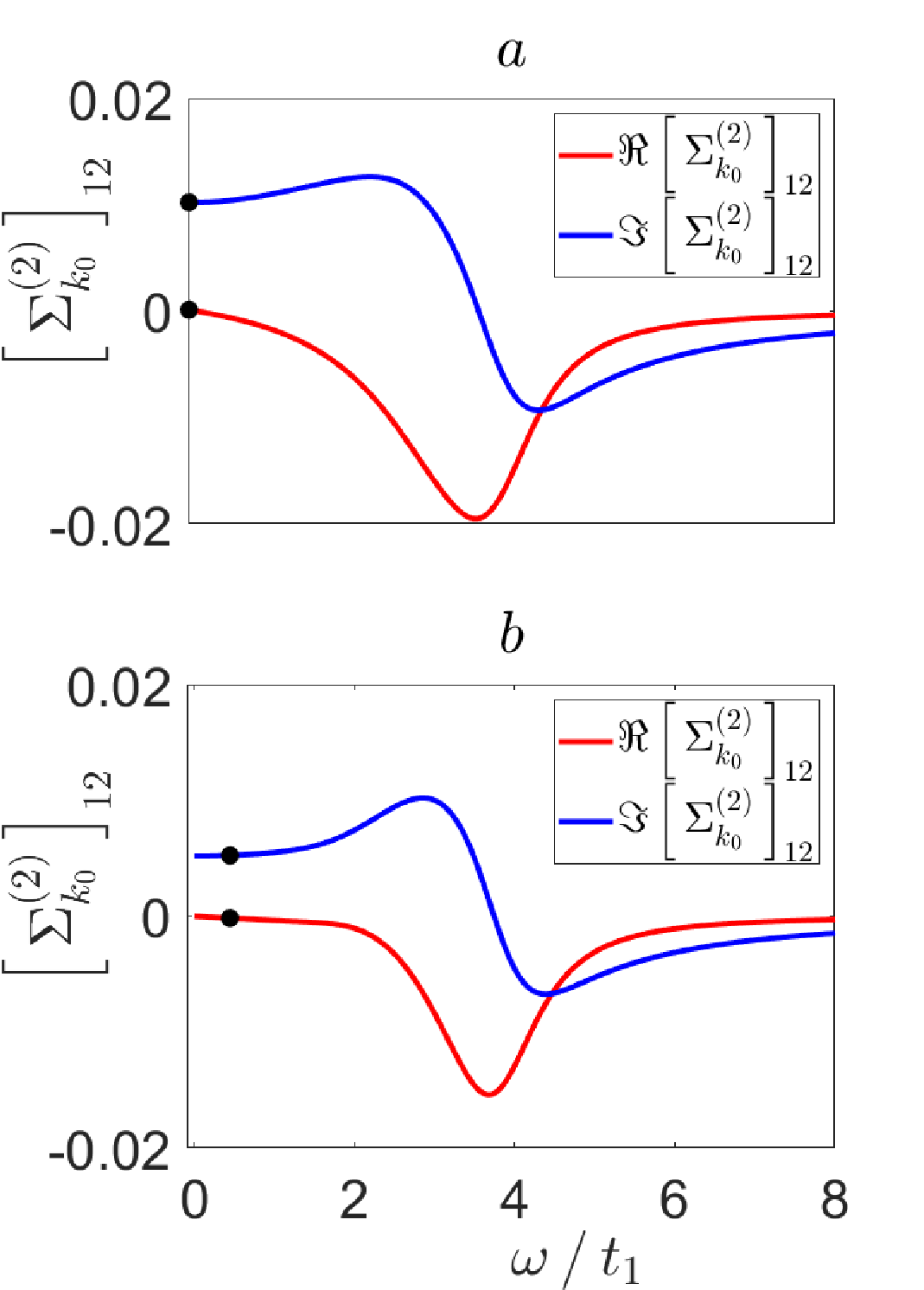}
\caption{Real and imaginary parts of anomalous self-energy $\left[\,\Sigma^{(2)}_{k_0}\,\right]_{12}$ as functions of frequency $\omega$ at the wave vector $k_0$ where the mean-field spectrum has minimum for parameters corresponding to the central point (a) and asterisk (b) on the Fig.\,\ref{9}a. The dots mark the functions at the %target 
mass shell, $\omega=\omega_k$. %frequency. 
It can be seen that $\im\left[\,\Sigma^{(2)}_{k_0}(0)\,\right]_{12} \cong \im\left[\,\Sigma^{(2)}_{k_0}(\omega_k)\,\right]_{12}$ and $\re\left[\,\Sigma^{(2)}_{k_0}(\omega_k)\,\right]_{12} \cong 0$ justifying the relations (\ref{X_X_bar_compare}) and (\ref{enk_denk}).}\label{6}
\end{figure}

Here we analyze the qualitative reasons why the nodal points of $Y_k$ can appear near the points of mean-field-spectrum minimum $k_0$.  
Let us start with the introduction of functions
\begin{eqnarray}\label{XY_bar}
&&X_{k} \to \bar{X}_{k} = \omega_{k}-\varepsilon_{k} - \re\left[\,\hat{\Sigma}_k(\omega_{k} + i0^+)\,\right]_{11},\nonumber\\
&&Y_{k} \to \bar{Y}_{k} = \im\left[\,\hat{\Sigma}_k(\omega_{k}+i0^+)\,\right]_{12},
\end{eqnarray}
that, in fact, define the real parts of the normal and anomalous components of the inverse Green's function at the %target 
frequency $\omega=\omega_{k} \in \mathbb{R}$. %, $\eta \to 0$.
The frequency $\omega_k$ is determined from the equation 
$$\re\left(\det\left[\,\hat{g}^{-1}_{k}(\omega_{k})- \hat{\Sigma}_{k}(\omega_{k})\,\right]\right)=0,$$
which is similar to Eq.\,(\ref{gamma_denk}), but takes into account the anomalous self-energy components. Bearing in mind that $|\gamma_{k_0}|\ll|\gamma_{\tilde{k}}|$, $|\delta\varepsilon_{k_0}|\ll|\delta\varepsilon_{k_{\tilde{k}}}|$, where $\tilde{k}$: $|\gamma_{\tilde{k}}|=\max\left(|\gamma_{k}|\right)$ (see Figs. \ref{8}c and \ref{7}a in  Sec. \ref{sec.3.2}), and the behavior of the anomalous self-energy shown in Fig. \ref{6}, the two branches of the renormalized spectrum can be approximately defined as 
\begin{align}
\bar{\varepsilon}_{k_0} & \approx \pm\sqrt{\bar{X}_{k_0}^2 + \bar{Y}_{k_0}^2}\,,\notag \\
\dot{\bar{\varepsilon}}_{k_0} & \sim \left(\,\bar{X}_{k_0}\,\dot{\bar{X}}_{k_0} + \bar{Y}_{k_0}\,\dot{\bar{Y}}_{k_0}\,\right)=0
\label{enk_denk}
\end{align}
and
\begin{eqnarray}\label{X_X_bar_compare}
X_{k_0} \approx \,\bar{X}_{k_0}\,,~~
Y_{k_0} \approx \,\bar{Y}_{k_0}.\,
\end{eqnarray}

Since there is the proximity-induced superconductivity in the system, it can be assumed that the functions $\bar{X}_{k}$ and $\bar{Y}_{k}$ are independent from each other. Therefore, to satisfy Eq.\,(\ref{enk_denk}) one has to require that
\begin{eqnarray}\label{XY_zero}
\bar{X}_{k_0}\,\dot{\bar{X}}_{k_0} = 0\,,~~\bar{Y}_{k_0}\,\dot{\bar{Y}}_{k_0}=0. 
\end{eqnarray}

As noted in Sec.\,\ref{sec.3.1}, the renormalized energy of the particle-like excitations is  $\bar{\varepsilon}_{k_0} \cong - \bar{X}_{k_0} \neq 0$, since the contributions from $\bar{Y}_{k_0}^2$ give the corrections $\sim \left(V\cdot \Gamma\right)^4$.
Therefore, $\dot{\bar{X}}_{k_0} \cong 0$ at the extreme point of the mean-filed spectrum. 

On the other hand, the derivatives $\dot{\bar{X}}_{k_0}$ and $\dot{\bar{Y}}_{k_0}$ cannot simultaneously become zero due to the condition (\ref{X_X_bar_compare}) and the fact that the vector $\left(\,\dot X_{k}\,,\,\dot Y_{k}\,\right)$ is a tangent to the fluctuation loop. Then, in order to satisfy Eq.\,(\ref{enk_denk}) and Eq.\,(\ref{XY_zero}), we should expect that
\begin{eqnarray}\label{XY_w0_cond}
\dot{\bar{X}}_{k_0} \cong 0\,,~~\bar{Y}_{k_0} \cong 0, ~~\dot{\bar{Y}}_{k_0} \neq 0,
\end{eqnarray}

Comparing expressions (\ref{X_X_bar_compare}) and (\ref{XY_w0_cond}), we see that the static part of the anomalous self-energy component $Y_k$, see Eq.\,(\ref{Xk_Yk_def}), should have the nodal points in the vicinity of the minimum of the mean-field spectrum $\varepsilon_k$. This is confirmed by the numerical calculations of the $k$-dependence of $\varepsilon_k$ %vs. $k$ 
and $Y_k$ % vs. $k$ dependencies 
for a wide range of the system parameters including the system with the long-range hoppings and superconducting pairings.

\bibliography{Majorana}% Produces the bibliography via BibTeX.

\end{document}